\documentclass[lettersize,journal]{IEEEtran}
\usepackage{amsmath,amsfonts}
\usepackage{algorithmic}
\usepackage{algorithm}

\usepackage{cite}

\usepackage{array}
\usepackage{textcomp}
\usepackage{stfloats}
\usepackage{url}
\usepackage{verbatim}
\usepackage{graphicx}
\usepackage{multirow}
\usepackage{hyperref}
\usepackage{subcaption}
\usepackage{xcolor}
\usepackage{amsthm}
\newcommand{\thicktilde}[1]{\mathbf{\tilde{\text{$#1$}}}}
\newcommand{\thickhat}[1]{\mathbf{\hat{\text{$#1$}}}}

\newcommand\numberthis{\addtocounter{equation}{1}\tag{\theequation}}

\begin{document}

\title{Of All StrIPEs: Investigating Structure-informed Positional Encoding for Efficient Music Generation}

\author{Manvi Agarwal,~\IEEEmembership{IEEE student member,}
Changhong Wang,~\IEEEmembership{IEEE member,} 
Ga\"{e}l Richard,~\IEEEmembership{IEEE Fellow.}

\thanks{
Preprint. Manuscript submitted for review.

All authors are based at Information Processing and Communications Laboratory (LTCI), T\'{e}l\'{e}com Paris, Institut Polytechnique de Paris, France. 

This work was funded by the European Union (ERC, HI-Audio, 101052978). Views and opinions expressed are however those of the author(s) only and do not necessarily reflect those of the European Union or the European Research Council. Neither the European Union nor the granting authority can be held responsible for them.}
}



\maketitle

\begin{abstract}
While music remains a challenging domain for generative models like Transformers, a two-pronged approach has recently proved successful: inserting musically-relevant structural information into the positional encoding (PE) module and using kernel approximation techniques based on Random Fourier Features (RFF) to lower the computational cost from quadratic to linear. Yet, it is not clear how such RFF-based efficient PEs compare with those based on rotation matrices, such as Rotary Positional Encoding (RoPE). In this paper, we present a unified framework based on kernel methods to analyze both families of efficient PEs. We use this framework to develop a novel PE method called \textit{RoPEPool}, capable of extracting causal relationships from temporal sequences. Using RFF-based PEs and rotation-based PEs, we demonstrate how seemingly disparate PEs can be jointly studied by considering the content-context interactions they induce. For empirical validation, we use a symbolic music generation task, namely, melody harmonization. We show that RoPEPool, combined with highly-informative structural priors, outperforms all methods.
\end{abstract}


\section{Introduction}

Generative models have been attracting increasing interest due to their impressive capability to produce high-quality, realistic samples. Among the ingredients necessary for obtaining this superior performance, two play a key role: voluminous data and ever-increasing parameter counts~\cite{kaplan_scaling_2020}.
Domains with large-scale datasets, such as vision, text and speech, use these ingredients to obtain effective large models. In contrast, since music lacks large publicly-available datasets, it is necessary to find new ways to efficiently use existing small-scale datasets.

Another important ingredient is recent architectural advances, such as the development of Transformers and attention.
Music has been no stranger to this phenomenon. However, music generated by such models often lacks essential properties that characterize human-generated `real' music, such as long-term coherence and multi-resolution organization~\cite{wu_jazz_2020}.
One simple way of improving music generation is to embed prior knowledge about musical structure into data-driven Transformers~\cite{ji_survey_2023,richard_model_2024,bhandari_motifs_2024} through the positional encoding (PE) module~\cite{agarwal_structure_2024,yi_popmag_2020,guo_domain_2023,liu2022symphony}, which has proven to be a good way to replace the usual (structure-free) PE without an increase in complexity. 

Despite their success, Transformers have a bottleneck: their complexity scales quadratically with sequence length. This can be reduced to a linear cost by using kernel approximations~\cite{tay_efficient_2022, tsai_transformer_2019}. 
Recently, low-complexity Transformers were combined with musically-relevant structural information with F-StrIPE~\cite{agarwal_fast_2025}, which generalizes an older, structure-free PE method, Stochastic Positional Encoding~\cite{liutkus_relative_2021}, by using Random Fourier Features (RFF)~\cite{rahimi_random_2007}.
However, an efficient PE that is conspicuously missing as a benchmark here is Rotary Positional Encoding (RoPE)~\cite{su_roformer_2024}. Conversely, works that investigate RoPE overlook RFF-based efficient PEs, though Stochastic Positional Encoding (SPE) was the first linear-complexity relative positional encoding.

\begin{figure}[!t]
\centering
\includegraphics[width=2.9in]{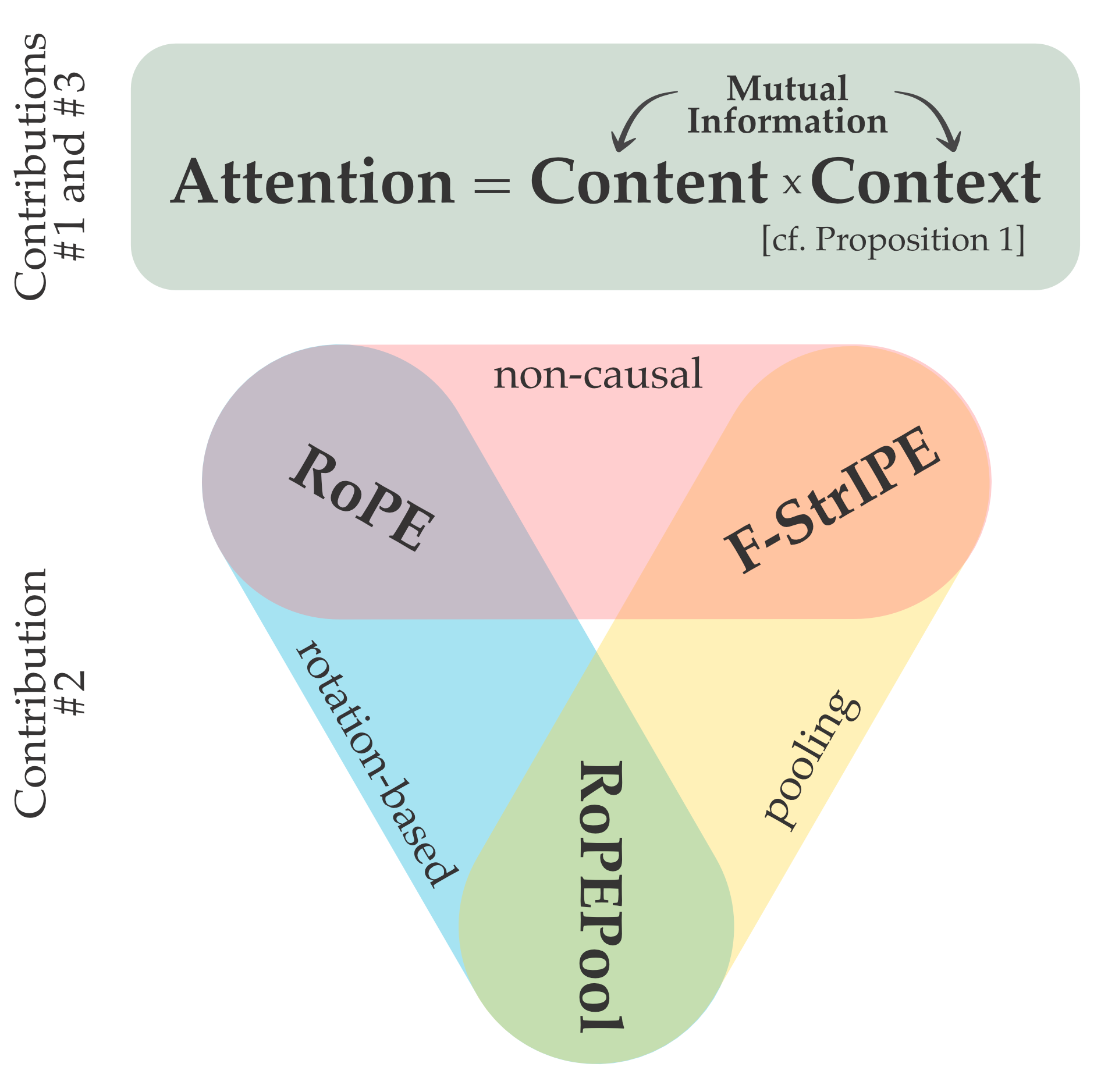}
\caption{Top: Efficient attention enriched with positional information can be viewed as a Tensor Product Kernel operating on context similarity and content similarity. Bottom: Relationship of the three positional encoding methods discussed in this paper (RoPE, F-StrIPE and RoPEPool)}
\label{fig:summary}
\end{figure}

This reveals a gap in the understanding of efficient positional encoding: how can we comparatively analyze efficient PEs that are based on Random Fourier Features, such as SPE and F-StrIPE, with efficient PEs that are based on rotation matrices, such as RoPE and its variants? A cursory reading of the literature reveals that most works compare PEs largely on a task-dependent basis, with some notable exceptions that also provide insights into the characteristics of different PE methods~\cite{ke_rethinking_2021, chang_convolutions_2021, chen_simple_2021}. We follow the latter line of thought, comparing PEs not just empirically through task performance, but also from a theoretical perspective, using \textit{kernel methods}.

We present three contributions in this paper, which are summarized in Fig. \ref{fig:summary}.

\textit{First}, we unify RFF-based PEs and rotation-based PEs into a common framework based on a canonical form of PE-enriched linear attention. We characterize this canonical form as a tensor product kernel operating on two spaces: one encoding similarity in terms of content and the other encoding similarity in terms of context. By studying the context similarity kernels produced by RoPE and F-StrIPE, we make it possible to compare the forms of attention they induce.

\textit{Second}, we introduce a novel offshoot of RoPE, called RoPEPool, which, unlike RoPE and F-StrIPE, is capable of extracting causal relationships from temporal data. We do this by framing PE-enriched attention as feature transforms applied to queries and keys and outline the feature transforms corresponding to RoPE and F-StrIPE. We find that F-StrIPE relies on the cross-dimension independence property to use pooling on its feature transform, while RoPE approximates attention by using its feature transform without pooling. We use this insight to combine RoPE's feature transform with the pooling operation to induce cross-dimension interactions, yielding RoPEPool. We use a synthetic dataset to illustrate the rich relationship between content similarity and context similarity in RoPEPool.

\textit{Third}, we provide new insights to explain why rich contextual information in positional encoding leads to better performance. We probe this idea empirically with an application to symbolic music generation. In particular, we use the melody harmonization task to comparatively assess the performance of RoPE, F-StrIPE and RoPEPool on different types of contextual information. We find that RoPEPool with highly-informative structural priors performs the best on our chosen task. We make our notion of the content-context interaction precise by computing the mutual information between the content and the context under different types of contextual information. We show that performance correlates strongly with mutual information and we provide a phenomenological analysis of how the performance of different PE methods changes as context carries more information about content.

The rest of this paper is organized as follows. In Section \ref{section:background}, we explain the background literature needed to set the stage for our work. Section \ref{section:methods}shows how different PEs can be studied using a unified framework of PE-enriched attention as well as present and analyze our novel PE method, RoPEPool. In Section \ref{section:experiments}, we introduce the experimental protocols we use for empirically validating the ideas we present in Section \ref{section:methods}. In Sections \ref{section:results} and \ref{section:discussion}, we present the results of our empirical enquiry and show that the concepts developed in Section \ref{section:methods} are valid for real datasets.

\section{Background} \label{section:background}

In the rest of the paper, we use $x$ for scalars, $\mathbf{x}$ for vectors and $\mathbf{X}$ for matrices. Further, we use the convention ``$\mathbf{Q}/\mathbf{K}$" to mean ``$\mathbf{Q}$ respectively $\mathbf{K}$''.

\subsection{Transformers, Attention and Positional Encoding}

The Transformer~\cite{vaswani_attention_2017} is a sequence-to-sequence architecture. Unlike recurrent neural networks, which perform sequential processing, Transformers process all timesteps in a sequence parallely using a mechanism called \textit{attention}. Given a sequence $[\mathbf{x}_1 , ... , \mathbf{x}_T]$ as input, each element of the output sequence is computed as:
\begin{align}
    \mathbf{y}_m = \frac{\sum_n \mathsf{a}_{m n} \mathbf{v}_n}{\sum_n \mathsf{a}_{m n}} \text{ with } \mathsf{a}_{mn} &= \text{exp} \bigg( \frac{a_{mn}}{\sqrt{D}} \bigg) \\
    a_{mn} &= \mathbf{q}_m \mathbf{k}_n^\top
\end{align}
where $a_{mn}$ is the attention coefficient or ``similarity score" for a pair of timesteps $(m,n)$.
Inputs $\mathbf{x}_m$ are transformed with linear layers to obtain queries $\mathbf{q}_m$, keys $\mathbf{k}_m$ and values $\mathbf{v}_m$. 
Because attention does parallel processing, it lacks the notion of causality. 
This makes Transformers invariant to permutations in the temporal order of inputs. Hence, \textit{positional encoding} (PE) is used to provide the model with a sense of time. Positional information can be included in two places: at the input (i.e., via $\mathbf{x}_m$) or during attention computation (i.e., via $a_{mn}$). Here, we focus on the latter approach, which is called Relative Positional Encoding (RPE).

\subsection{Efficient Attention with Positional Information} \label{ssection:spe}

The computational cost of attention scales quadratically with sequence length. To bring this cost down to the linear regime, kernelized attention was proposed:
\begin{equation} \label{eq:attention_approx}
    \mathsf{a}_{mn} = \mathcal{K} ( \mathbf{q}_m , \mathbf{k}_n ) = \mathbb{E} \Big[ \phi(\mathbf{q}_m) \phi(\mathbf{k}_n)^{\top} \Big]
\end{equation}
where $\mathcal{K}$ is a positive definite kernel and $\phi(\mathbf{x}): \mathbb{R}^D \to \mathbb{R}^{D_\phi}$ defines a \textit{randomized feature map} for $\mathbf{x}$~\cite{tsai_transformer_2019, choromanski_rethinking_2021, katharopoulos_transformers_2020, zhuoran_efficient_2021}. Since it uses multiple instantiations, $\phi$ captures, on average, the relationship between $\mathbf{q}_m$ and $\mathbf{k}_n$, typified by $\mathcal{K}$. As a result, coefficients $\mathsf{a}_{mn}$ do not need to be computed explicitly, which produces linear-complexity Transformers.

This efficient, kernelized formulation cannot be directly used with RPE which, as introduced in \cite{shaw_rpe_2018}, requires the explicit computation of attention coefficients $a_{mn}$. 
This lacuna was filled by Stochastic Positional Encoding~\cite{liutkus_relative_2021}, which, uses the following \textit{canonical form of PE-enriched exact attention}:
\begin{equation} \label{eq:attention_canonical_form}
    a_{mn} = \sum_{d=1}^D a^d_{mn} = \left[ \sum_{d=1}^D \text{diag}(\mathbf{Q}_{:,d}) \mathbf{P}_d \text{diag}(\mathbf{K}_{:,d}) \right]_{mn}
\end{equation}
The positional matrix $\mathbf{P}_d$ captures the relationship between all pairs $(m, n)$ of timesteps that come from the positional index sequences $\mathcal{P}_Q = \{ 1, ..., m, ..., T_Q \}$, associated with the query matrix $\mathbf{Q}$, and $\mathcal{P}_K = \{ 1, ..., n, ..., T_K \}$, associated with the key matrix $\mathbf{K}$.
In this equation, $\mathbf{Q}_{:, d}/\mathbf{K}_{:, d}$ extracts a $T_Q/T_K$-dimensional vector containing the $d^{th}$ dimension for all timesteps of the query/key matrix. To obtain approximate attention, SPE factorizes $\mathbf{P}_d$ as:
\begin{equation} \label{eq:spe:ohyeah}
a_{mn} \approx \left[ \sum_{d=1}^D
    \lefteqn{\overbrace{\phantom{
    \text{diag}(\mathbf{Q}_{:,d})
    \frac{
        \thicktilde{\mathbf{P}}^Q_d
    }{\sqrt{R}}
    }}^{
    \mathbf{Q}^{\text{SFF}}
    }}
    \text{diag}(\mathbf{Q}_{:,d})
    \underbrace{
    \frac{
        \thicktilde{\mathbf{P}}^Q_d
    }{\sqrt{R}}
    \lefteqn{\overbrace{\phantom{
    \frac{
        \thicktilde{\mathbf{P}}^K_d{}^\top
    }{\sqrt{R}}
    \text{diag}(\mathbf{K}^\top_{:,d})
    }}^{
    \mathbf{K}^{\text{SFF}}
    }}
    \frac{
        \thicktilde{\mathbf{P}}^K_d{}^\top
    }{\sqrt{R}}
    }_{
    \approx \mathbf{P}_d
    }
    \text{diag}(\mathbf{K}^\top_{:,d})
    \right]_{mn}
\end{equation}
\begin{equation} \label{eq:sff}
    \thicktilde{\mathbf{P}}^{Q/K}_d = \frac{\boldsymbol{\Omega}\left(\mathcal{P}_{Q/K}, \boldsymbol{f}_d, \boldsymbol{\theta}_d^{Q/K}\right) \text{diag}\left(\ddot{\boldsymbol{\lambda}_d}\right) \mathbf{Z}_d}{\sqrt{2 N_f}}
\end{equation}
The matrices $\mathbf{Q}^{\text{SFF}}$ and $\mathbf{K}^{\text{SFF}}$ combine \textit{content} information ($\mathbf{Q}$/$\mathbf{K}$) with \textit{context} information ($\thicktilde{\mathbf{P}}^Q_d$/$\thicktilde{\mathbf{P}}^K_d$), using Stochastic Fourier Features (SFF), and can be seen as PE-enriched versions of $\mathbf{Q}$ and $\mathbf{K}$. 
The matrices $\thicktilde{\mathbf{P}}^Q_d$ and $\thicktilde{\mathbf{P}}^K_d$ contain representations of the positional indices associated with the queries $\mathbf{Q}$ and the keys $\mathbf{K}$.
As shown in (\ref{eq:sff}), these matrices 
collect $R$ feature realizations of index sequences $\mathcal{P}_Q$ and $\mathcal{P}_K$, which provide an approximation of the positional matrix $\mathbf{P}_d$.
More precisely, in (\ref{eq:sff}), the matrix $\boldsymbol{\Omega}\left(\mathcal{P}_{Q/K}, \boldsymbol{f}_d, \boldsymbol{\theta}_d^{Q/K}\right)$ contains sinusoidal features for the index sequence $\mathcal{P}_{Q/K}$, parameterized by $N_f$ frequencies (collected in $\boldsymbol{f}_d$) and phase shifts (collected in $\boldsymbol{\theta}_d$). The diagonal matrix $\text{diag}\left(\ddot{\boldsymbol{\lambda}_d}\right)$ is used to apply gains to the sinusoidal features. The matrix $\mathbf{Z}_d$, consists of i.i.d. entries from a zero-mean, unit-variance Gaussian distribution.
SPE goes one step beyond (\ref{eq:spe:ohyeah}), by modifying it to:
\begin{equation} \label{eq:attention_spe_approx_form}
a_{mn} \approx \left[ \Bigg( 
    \underbrace{
    \sum_{d=1}^D
    \text{diag}(\mathbf{Q}_{:,d})
    \frac{
        \thicktilde{\mathbf{P}}^Q_d
    }{\sqrt{R}}
    }_{\mathbf{Q}^{\text{SFF}}_c}
    \Bigg) \Bigg(
    \underbrace{
    \sum_{d=1}^D
    \text{diag}(\mathbf{K}_{:,d})
    \frac{
        \thicktilde{\mathbf{P}}^K_d{}
    }{\sqrt{R}}
    }_{\mathbf{K}^{\text{SFF}}_c}
    \Bigg)^\top
    \right]_{mn} \\
\end{equation}
The assumption made here, as noted earlier~\cite{liutkus_relative_2021}, is that the cross-dimension terms in (\ref{eq:spe:ohyeah}) are negligible, which allows the equivalence from (\ref{eq:spe:ohyeah}) to (\ref{eq:attention_spe_approx_form}) to hold. SPE applies the feature maps $\phi$, used to approximate attention (\ref{eq:attention_approx}), to $\mathbf{Q}^{\text{SFF}}_c$ and $\mathbf{K}^{\text{SFF}}_c$ instead of $\mathbf{Q}$ and $\mathbf{K}$. Hence, SPE applies a \textit{pooling} operation to the PE-enriched representations of $\mathbf{Q}$ and $\mathbf{K}$ \textit{before} applying the attention feature maps $\phi$ to them.




\subsection{Rotary Positional Encoding} \label{ssection:rope}

Parallel to the developments around SPE and F-StrIPE, another efficient positional encoding method, called Rotary Positional Encoding (RoPE)~\cite{su_roformer_2024}, has emerged in recent times.
As its name suggests, RoPE is based on rotation matrices, unlike SPE and F-StrIPE, which are based on RFF.
More precisely, RoPE approximates PE-enriched attention by applying rotation matrices $\thickhat{\mathbf{P}}^Q_m$ and $\thickhat{\mathbf{P}}^K_n$ to query and key vectors $\mathbf{q}_m$ and $\mathbf{k}_n$:
\begin{equation} \label{eq:rope_attention}
    a_{mn} \approx \mathbf{q}_m \big( \thickhat{\mathbf{P}}^Q_m \big)^\top \thickhat{\mathbf{P}}^K_n \mathbf{k}_n^\top
\end{equation}
such that the rotations are parameterized by positions $m$ (for the query) and $n$ (for the key). The rotation matrices take a block diagonal form, given as:
\begin{equation}
    \thickhat{\mathbf{P}}^A_j
=\left[ \begin{array}{ccccc}
\mathbf{R}_j^A (1) & \cdots & 0 \\
\vdots & \ddots & \vdots \\
0 & \cdots & \mathbf{R}_j^A (\frac{D}{2})
\end{array}\right]
\end{equation}
where each $\mathbf{R}$ is a $2 \times 2$ rotation matrix, given as:
\begin{equation}\label{eq:rotation_matrix}
\thickhat{\mathbf{R}}^A_j (i)
=\left[ \begin{array}{ccccc}
\cos (f_i \mathcal{P}_A[j] ) & -\sin (f_i \mathcal{P}_A[j] ) \\
\sin (f_i \mathcal{P}_A[j] ) & \cos (f_i \mathcal{P}_A[j] )
\end{array}\right]
\end{equation}

As hinted in the introduction, RoPE has become tremendously popular, having been deployed in several Large Language Models (LLMs), such as GPT-NeoX~\cite{black_gptneox_2022}, Gemma~\cite{gemma_gemma_2024}, the LLaMa model family~\cite{touvron_llama_2023} and the PaLM model family~\cite{chowdhery_palm_2023}.
With its integration into the PyTorch ecosystem, for example, it has become the PE method of choice for machine learning practitioners working with Transformers.
Several works have also dissected and improved on the properties of RoPE, such as its length generalization abilities~\cite{kaikondev_things_2023, chen_extending_2023, wang_resonance_2024, pawar_what_2024}. This line of work has lead to insights on its advantages, particularly from a mechanistic interpretability perspective~\cite{barbero_round_2025}.

\section{Methods} \label{section:methods}

\subsection{F-StrIPE: Fast, Structure-informed Positional Encoding} \label{ssection:spe-fstripe}

SPE can be directly connected to kernel methods by noting that SPE is a noisy version of Random Fourier Features (RFF)~\cite{rahimi_random_2007}. The noise-free version of SPE, which directly uses RFF to encode musically-relevant structural information in positional encoding, is called F-StrIPE~\cite{agarwal_fast_2025}.

F-StrIPE reformulates SPE by making two changes. First, it reconceptualizes positional encoding as a flexible way to introduce domain-specific prior knowledge about the underlying data domain of interest. As can be seen in SPE (Section \ref{ssection:spe}), positional encoding depends on a sequence $\mathcal{P} = [ p_1, p_2, ... , p_T ]$ of positional indices. Instead of using time information, with $p_i = i$, as is usually done in standard PE, F-StrIPE allows the use of multi-resolution structural labels, with $p_i = \mathbf{s}(i)$, where $\mathbf{s}(i)$ is a vector with $L$ structural labels. Second, it modifies $\thicktilde{\mathbf{P}}^{Q/K}_d$ by noting that the term $\mathbf{Z}_d \mathbf{Z}_d^\top / R = \widehat{\mathbf{C}}_d$ appears in the approximation of $\mathbf{P}_d$ by using Equations (\ref{eq:spe:ohyeah}) and (\ref{eq:sff}). $\widehat{\mathbf{C}}_d$ acts as an empirical covariance matrix and its elements can be written as random variables, with $\alpha = \frac{1}{R} \sum_{i=1}^{R} \upsilon_i^2$ on the diagonal and $\beta = \frac{1}{R} \sum_{i=1}^{R} \upsilon_i \nu_i$ on the off-diagonals, where $\upsilon_i$ and $\nu_i$ are standard Gaussian random variables. 
F-StrIPE operates in the limit of $R \to \infty$, where the values of $\alpha$ and $\beta$ converge to $\alpha \to 1$ and $\beta \to 0$ and, as a consequence, the empirical covariance matrix $\widehat{\mathbf{C}}_d$ converges to the theoretical covariance matrix $\mathbf{C}_d = \mathbf{I}_{2N_{f}}$.
Thus, F-StrIPE takes the ideal case of $\mathbf{C}_d$, where the approximation of $\mathbf{P}_d$ simplifies to:
\begin{equation} \label{eq:ideal_C}
    \mathbf{P}_d[m, n] \approx  \frac{1}{N_f}  \sum_{\omega = 1}^{N_f} \Lambda_\omega \cos \big( f_{\omega} ( \mathcal{P}_Q[m] - \mathcal{P}_K[n] ) + \Theta_\omega \big)
\end{equation}
where $\Lambda_\omega$ is the gain contributed by the matrices $\text{diag}\left(\ddot{\boldsymbol{\lambda}_d}\right)$ and $\Theta_\omega$ is the phase-shift contributed by $\boldsymbol{\theta}^Q_d$ and $\boldsymbol{\theta}^K_d$. By adjusting the $\frac{1}{2}$ scaling factor, coming from (\ref{eq:sff}), F-StrIPE obtains an expression that contains a mean. Thus, F-StrIPE results in an approximation of $\mathbf{P}_d$ (\ref{eq:ideal_C}) that is precisely the form used by RFF for approximating stationary, positive definite kernels, yielding a direct connection between positional encoding and kernel approximation.
Using this insight, F-StrIPE redesigns the positional feature matrices of (\ref{eq:sff}) as:
\begin{equation} \label{eq:rff}
    \mathbf{P}^{Q/K}_d = \boldsymbol{\Omega}\left(\mathcal{P}_{Q/K}, \boldsymbol{f}_d, \boldsymbol{\theta}_d^{Q/K}\right) \text{diag}\left(\ddot{\boldsymbol{\lambda}_d}\right) / \sqrt{N_f}
\end{equation}
where the sinusoidal feature matrix $\boldsymbol{\Omega}$ uses structure-aware positional indices as described earlier in this section. 
With this, F-StrIPE can replace $\thicktilde{\mathbf{P}}^{Q/K}_d$ by $\mathbf{P}^{Q/K}_d$, yielding $\mathbf{Q}^{\text{RFF}}/\mathbf{K}^{\text{RFF}}$ in place of $\mathbf{Q}^{\text{SFF}}/\mathbf{K}^{\text{SFF}}$ for (\ref{eq:spe:ohyeah}) and $\mathbf{Q}^{\text{RFF}}_c/\mathbf{K}^{\text{RFF}}_c$ in place of $\mathbf{Q}^{\text{SFF}}_c/\mathbf{K}^{\text{SFF}}_c$ for (\ref{eq:attention_spe_approx_form}).
F-StrIPE inherits the pooling technique from SPE (\ref{eq:attention_spe_approx_form}), leading to the attention feature maps $\phi$ being applied to $\mathbf{Q}^{\text{RFF}}_c$ and $\mathbf{K}^{\text{RFF}}_c$ in place of $\mathbf{Q}$ and $\mathbf{K}$.

\subsection{Interpreting PE-enriched Attention through Kernels}

At first glance, F-StrIPE and RoPE seem to be completely divergent forms of PE-enriched attention. However, as we will explain in the following sections, they can be jointly analyzed with kernel methods. It is indeed possible to reframe PE-enriched attention in the idiom of kernels by considering the canonical form of PE-enriched exact attention (\ref{eq:attention_canonical_form}) presented in Section \ref{ssection:spe}, which gives us our first insight:

\vspace{2mm}
\noindent {\bf{Proposition 1:}} Exact attention enriched with positional information (\ref{eq:attention_canonical_form}) can be written as a Tensor Product Kernel~\cite{scholkopf_learning_2001}:
\begin{equation} \label{eq:tensor_product_kernel}
    a_{mn}^d = \mathcal{F}(q_{md}, k_{nd}) \times \mathcal{G}_d (p_m, p_n)
\end{equation}
where $\mathcal{F}$ is the linear kernel $\mathcal{F}(x, x^\prime) = x \cdot x^\prime $ applied to $q_{md}$ and $k_{nd}$, capturing \textit{content similarity}, and $\mathcal{G}_d$ is a kernel that depends on the form of $\mathbf{P}_d$ and is applied to positional indices $p_m$ and $p_n$, capturing \textit{context similarity}.
Changing the form and approximation of $\mathbf{P}_d$ (\ref{eq:attention_canonical_form}) and, consequently $\mathcal{G}_d$ (\ref{eq:tensor_product_kernel}) changes the specific form of PE-enriched attention we admit. Thus, we can infer the properties of $a_{mn}$ by decomposing it into its constituents $a^d_{mn}$ and analyzing their properties, specifically by verifying the forms that $\mathcal{F}$ and $\mathcal{G}_d$ take.
We provide a sketch proof of Proposition 1 in the supplementary material.

\subsection{Comparing Rotation-based and RFF-based PEs}

In this section, we will describe and compare RoPE and F-StrIPE in two ways.
First, we will characterize the nature of attention ($a_{mn}$) induced by each of these PE methods by studying the form of $\mathcal{G}_d$ (Proposition 1) they use. Second, we will contrast these methods by reinterpreting PE-enriched attention as feature transforms applied to queries and keys.

For our first goal, we start with a characterization of RoPE through the lens of kernels. We can derive an expression using (\ref{eq:rope_attention})  and (\ref{eq:rotation_matrix} for RoPE, which is as follows:
\begin{align*}
    a_{mn} \approx \sum_{d=1}^{D/2} 
    & \underbrace{(q_{m,2d-1} k_{n,2d-1} + q_{m,2d} k_{n,2d} )}_{\mathcal{F}_1}
    \underbrace{\cos \Delta^-_d}_{\mathcal{G}_d^{(1)}} \hspace{2mm} + \\ &
    \underbrace{( q_{m,2d-1} k_{n,2d} - q_{m,2d} k_{n,2d-1} )}_{\mathcal{F}_2} 
    \underbrace{\sin \Delta^-_d}_{\mathcal{G}_d^{(2)}}
    \numberthis \label{eq:rope_attention_approx}
\end{align*}
where we used \textit{pairs of dimensions} as our unit of analysis and we set $\Delta^-_d = f_d ( \mathcal{P}_Q[m] - \mathcal{P}_K[n] )$.
We can see that RoPE captures content similarity using $\mathcal{F}_1$ and $\mathcal{F}_2$ and context similarity using $\mathcal{G}_d^{(1)}$ and $\mathcal{G}_d^{(2)}$. Since $\mathcal{F}_2$ is not symmetric in its arguments, $\mathcal{F}_2$ is not a positive definite kernel. As a result, in the case of RoPE, $a^d_{mn}$ and, by extension, $a_{mn}$ are also not positive definite kernels.

To compare this with F-StrIPE, we start by noting that the approximation of attention in F-StrIPE, using multiple frequencies for each dimension $d$ (\ref{eq:rff}), can be written as $a_{mn} \approx \sum_{d=1}^D q_{md} k_{nd} \mathbf{P}_d [m, n]$, 
where we used $\mathbf{P}_d[m,n]$ from (\ref{eq:ideal_C}).
Thus, the role of RFF is to obtain an approximation of $\mathbf{P}_d$, using $N_f$ frequencies for each dimension $d = 1, ..., D$, which is then applied to $q_{md} k_{nd}$.
In this way, multiple frequencies are used to approximate a PE-enriched version of the product $q_{md} k_{nd}$.
$N_f >> 1$ would be necessary and desirable if we used fixed parameters for the random features~\cite{rahimi_weighted_2008}. However, since F-StrIPE is in the regime of learnable parameters, we can directly learn the appropriate frequency, gain and offset for each dimension $d$. Thus, we trade-off learning-free randomization with learning-dependent optimization. In the case of learnable frequencies, with $N_f = 1$, we have:
\begin{equation} \label{eq:fstripe1_attention_approx}
    a_{mn} \approx \sum_{d=1}^D \underbrace{q_{md} k_{nd}}_{\mathcal{F}} \underbrace{\cos \Delta^-_d}_{\mathcal{G}_d}
\end{equation}
where we absorbed the learnable gains $\Lambda_d$ into the optimization of $q_{dm}$ and $k_{dn}$ and set the phase-shifts $\Theta_d$ to 0 in order to facilitate the comparison with RoPE. We call this method \textit{F-StrIPE$_1$} because it uses a single learnable frequency for each dimension. Since both $\mathcal{F}$ and $\mathcal{G}$ are positive definite kernels in this case, (\ref{eq:fstripe1_attention_approx}) gives us a positive definite kernel for $a_{mn}^d$ and $a_{mn}$. 

Note that, while a generalized framework of linearized relative positional encoding has been previously proposed~\cite{qin_linearized_2023}, it fails to make a link between PE-enriched attention and kernels. For Proposition 1, we used the convention of treating a dimension $d$ as our unit of analysis, but this can easily be extended to general units of analysis, such as \textit{collections of dimensions}, such that the sum in (\ref{eq:attention_canonical_form}) applies to such collections instead of individual dimensions. Indeed, as we saw, this is precisely what allows us to unify the analysis of RoPE and F-StrIPE.

We can further sharpen our understanding of the differences between rotation-based and RFF-based PEs by turning to our second goal and reframing PE-enriched attention as applying feature transforms to the query and key matrices. In this way, we can study the feature transforms induced by RoPE, on the one hand, and F-StrIPE$_1$, on the other.

We begin by returning to a fact we briefly mentioned in Section \ref{ssection:spe-fstripe}: F-StrIPE$_1$ inherits the pooling operation from SPE. The implication is that, in order for F-StrIPE$_1$ to use the form in (\ref{eq:attention_spe_approx_form}) for approximating attention, similar to SPE, the cross-dimension terms must be negligible. To understand whether this assumption is valid for F-StrIPE$_1$, we regard (\ref{eq:spe:ohyeah}) and (\ref{eq:attention_spe_approx_form}) as feature transforms. For (\ref{eq:spe:ohyeah}), we can write the attention matrix $\left[\mathbf{A}\right]_{mn} = a_{mn}$ as:
\begin{equation} \label{eq:unpooled_multiplication}
    \mathbf{A} \approx 
\begin{bmatrix}
\mathcal{U}(\mathbf{q}_1) \hspace{2mm}
\cdots \hspace{2mm}
\mathcal{U}(\mathbf{q}_{T_Q})
\end{bmatrix}^\top
\begin{bmatrix}
\mathcal{U}(\mathbf{k}_1) \hspace{2mm}
\cdots \hspace{2mm}
\mathcal{U}(\mathbf{k}_{T_K})
\end{bmatrix}
\end{equation}
where $\mathcal{U}(\cdot)$ is an \textit{unpooled feature transform}, applied to both queries and keys, which can be written as:
\begin{equation} \label{eq:unpooled_feature_map}
    \mathcal{U}(\mathbf{a}_i) : \mathbf{a}_i \mapsto \left[ h(a_{i1}) \oplus h(a_{i2}) \oplus \cdots \oplus h(a_{iD}) \right] 
\end{equation}
Here, we used $\mathbf{a}_i$ to stand in for $\mathbf{q}_m$ and $\mathbf{k}_n$ and $h$ as a function that is applied independently to each unit of analysis, which, in this case, is a single dimension. Thus, each dimension of each query/key vector is transformed using $h(\cdot)$ and concatenated together, with $\oplus$ being the concatenation operation, to obtain $\mathcal{U}(\cdot)$. The role of $h$ is to transform PE-free queries and keys into PE-enriched queries and keys, equivalent to the role played by $\mathcal{G}$ with respect to $\mathcal{F}$ (Proposition 1).
In contrast to (\ref{eq:spe:ohyeah}), for (\ref{eq:attention_spe_approx_form}), we can write the attention matrix $\mathbf{A}$ as:
\begin{equation} \label{eq:pooled_multiplication}
    \mathbf{A} \approx 
\begin{bmatrix}
\mathcal{C}(\mathbf{q}_1) \hspace{2mm}
\cdots \hspace{2mm}
\mathcal{C}(\mathbf{q}_{T_Q})
\end{bmatrix}
\begin{bmatrix}
\mathcal{C}(\mathbf{k}_1) \hspace{2mm}
\cdots  \hspace{2mm}
\mathcal{C}(\mathbf{k}_{T_K})
\end{bmatrix}^\top
\end{equation}
where $\mathcal{C}(\cdot)$ is a \textit{pooled feature transform}:
\begin{equation} \label{eq:pooled_feature_map}
    \mathcal{C}(\mathbf{a}_i) : \mathbf{a}_i \mapsto \left[ h(a_{i1}) + h(a_{i2}) + \cdots + h(a_{iD}) \right]
\end{equation}
where components are added instead of being concatenated.

We can see that $\mathcal{U}(\mathbf{q}_m) \mathcal{U}(\mathbf{k}_n)^\top = \mathcal{C}(\mathbf{q}_m) \mathcal{C}(\mathbf{k}_n)^\top$ only when the \textit{cross-dimension independence property} is fulfilled, i.e., $h(q_{md}) h(k_{nd^{\prime}}) \approx 0$ for $d \neq d^\prime$. In other words, this is precisely when (\ref{eq:spe:ohyeah}) and (\ref{eq:attention_spe_approx_form}) are equivalent.

Naturally, whether this property is met depends strongly on the nature of $h$. For F-StrIPE$_1$, $h$ is given by:
\begin{equation} \label{eq:fstripe1_h}
    h(a_{id}) = 
    \begin{bmatrix}
    a_{id} \cos (f_d p_i) & a_{id} \sin (f_d p_i)
    \end{bmatrix}
\end{equation}
With some simple ideas from harmonic analysis, we can show that \textit{the cross-dimension independence property holds for F-StrIPE$_1$}, making it possible to use (\ref{eq:attention_spe_approx_form}) with F-StrIPE$_1$. A sketch proof is provided in the supplementary materials.

Having described the feature transform for F-StrIPE$_1$, we turn to the feature transform corresponding to RoPE. This has already been described in the literature~\cite{su_roformer_2024} as:
\begin{align} 
    \mathcal{U}(\mathbf{a}_i) &: \mathbf{a}_i \mapsto \left[ h(a_{i(1:2)}) \oplus \cdots \oplus h(a_{i(D-1:D)}) \right] \\
    h(a_{i(d:d+1)}) &= \begin{bmatrix}
    a_{i,d} \cos (f_d p_i) - a_{i,d+1} \sin (f_d p_i) \\ 
    a_{i,d+1} \cos (f_d p_i) + a_{i,d} \sin (f_d p_i)
    \end{bmatrix}^\top \numberthis \label{eq:rope_g}
\end{align}
where $h$ is a function that is now applied independently to pairs of dimensions $(d, d+1)$. Note that $\mathcal{U}(\mathbf{q}_m) \mathcal{U}(\mathbf{k}_n)^\top$ gives us precisely the expected expression for $a_{mn}$ with RoPE (\ref{eq:rope_attention}).

At this point, it is pertinent to observe that SPE and, by extension, F-StrIPE$_1$ use pooling before applying $\phi$ (\ref{eq:attention_spe_approx_form}) to approximate downstream linear attention. However, it is possible to directly apply $\phi$ without pooling while maintaining linear computational complexity by using (\ref{eq:spe:ohyeah}) through its representation as an unpooled feature transform (\ref{eq:unpooled_multiplication}). The advantage of working directly with (\ref{eq:spe:ohyeah}) is that it can allow us to use functions $h$ where the cross-dimension independence property is not met. The drawback is that we introduce many more dimensions downstream for approximating linear attention with $\phi$. In fact, in contrast with F-StrIPE$_1$, which is able to use (\ref{eq:attention_spe_approx_form}) and a pooled feature transform for approximating attention, RoPE directly works with (\ref{eq:spe:ohyeah}) and an unpooled feature transform.




To summarize, we showed in this section that RoPE and F-StrIPE$_1$ can be comparatively understood via two perspectives:
\begin{enumerate}
    \item Using kernel methods: F-StrIPE$_1$ induces a positive definite kernel for attention, whereas RoPE induces a non-positive definite kernel.
    \item Rethinking attention as feature transforms: F-StrIPE$_1$ possesses the cross-dimension independence property, which allows pooling before applying $\phi$. On the other hand, RoPE does not have this property and directly uses $\phi$ without pooling.
\end{enumerate}
Perspective (2) also allows us to conclude that, in fact, it is not necessary to apply pooling in order to use downstream linear attention (\ref{eq:attention_spe_approx_form}). Rather, $\phi$ can directly be used without pooling (\ref{eq:spe:ohyeah}), as done in RoPE.

\subsection{RoPEPool: Gaining expressivity using simple operations}



Since RoPE does not possess the cross-dimension independence property, additional expressivity can be obtained through the cross-dimension interactions by combining RoPE's $h$ function (\ref{eq:rope_g}) with a pooled feature transform (\ref{eq:pooled_feature_map}) to compute $a_{mn}$. With this, we can explore how expressive functions $h$ can be leveraged to produce rich interactions between content and context by swapping concatenation for simple operations, such as addition.
We then obtain the following expression for attention, analogous to RoPE (\ref{eq:rope_attention_approx}) and F-StrIPE$_1$ (\ref{eq:fstripe1_attention_approx}):
\begin{align*}
    a_{mn} \approx \sum_{d=1}^{D/2} \Big\{ 
    &(q_{m,2d-1} k_{n,2d-1} + q_{m,2d} k_{n,2d}) \cos \Delta^- \hspace{1mm} + \\
    &(q_{m,2d-1} k_{n,2d} - q_{m,2d} k_{n,2d-1}) \sin \Delta^- \hspace{1mm} + \\
    &(q_{m,2d-1} k_{n,2d-1} - q_{m,2d} k_{n,2d}) \sin \Delta^+ \hspace{1mm} + \\
    &(q_{m,2d-1} k_{n,2d} + q_{m,2d} k_{n,2d-1}) \cos \Delta^+ \Big\} \numberthis \label{eq:ropeplus_attention_approx}
\end{align*}
where we used $\Delta^+ = f_{d} ( \mathcal{P}_Q[m] + \mathcal{P}_K[n] )$. 
Similar to our analysis for RoPE and F-StrIPE$_1$, the above equation induces a form of PE-enriched attention that is a non-pd kernel.
Further, in contrast to F-StrIPE$_1$ and RoPE, (\ref{eq:ropeplus_attention_approx}) depends not only on the lag between positional indices but also their sum. Thus, the cross-dimension interactions, obtained by using a pooled feature transform on RoPE's $h$, lead to a richer contextual representation in attention by introducing \textit{asymmetry}. Since this new method is obtained by a simple pooling operation on RoPE, we call it \textit{RoPEPool}.



So far, we have focused on distinguishing between different PE methods in rather formal terms, for example, by locating these methods within different kernel families. In the next section, we will combine this with an interpretive approach to compare their expressivities by, first, introducing some simplifications to (\ref{eq:rope_attention_approx}), (\ref{eq:fstripe1_attention_approx}) and (\ref{eq:ropeplus_attention_approx}) to make our comparisons clearer, and second, shedding light on the benefits of the asymmetrical context information used by RoPEPool.

\subsection{A Simplified Analysis of F-StrIPE$_1$, RoPE and RoPEPool} \label{ssection:toy_example}


To study how content and context interact under different PE methods, we start by making some mild assumptions about the query vectors $\mathbf{q}_m$ and key vectors $\mathbf{k}_n$.
We first assume that our basic unit of analysis is 2-dimensional chunks of the query and key vectors, given as $\mathbf{q}_{m,2d-1:2d}$ and $\mathbf{k}_{n,2d-1:2d}$. This does not change anything for RoPE and RoPEPool, but it unifies pairs of dimensions in F-StrIPE$_1$ (\ref{eq:fstripe1_attention_approx}) under a single approximation of $\mathcal{G}_d$. Further, we assume that each 2-dimensional chunk has a Euclidean norm of 1.
Then, since we can write the coordinates of any 2D unit-norm vector $\mathbf{v}$ as $\left[ \cos \psi_v , \sin \psi_v \right]$, where $\psi_v$ is the angle that $\mathbf{v}$ makes with the X-axis, we can think of these chunks solely in terms of their angular representations.

With this, we can first rewrite (\ref{eq:rope_attention_approx}) for RoPE as:
\begin{equation} \label{eq:rope_toy_equation}
    a_{mn} \approx \sum_{d=1}^{D/2} \cos \big( ( \psi_q^d - \psi_k^d ) + f_d ( \xi_m^d - \xi_n^d ) \big)
\end{equation}
Here, $\psi_q^d$ and $\psi_k^d$ are the angles made by $\mathbf{q}_{m,2d-1:2d}$ and $\mathbf{k}_{n,2d-1:2d}$ with the X-axis, whereas $\xi_m^d$ and $\xi_n^d$ are the angular representations of $\mathcal{P}_Q[m]$ and $\mathcal{P}_K[n]$ in radians. The frequency parameter $f_d$ varies between 0 and 1 and controls the contribution of contextual information to attention.

Similarly, we can rewrite (\ref{eq:fstripe1_attention_approx}) for F-StrIPE$_1$ as:
\begin{equation} \label{eq:fstripe1_toy_equation}
    a_{mn} \approx \sum_{d=1}^{D/2} \cos \big( \psi_q^d - \psi_k^d \big) \cos \big( f_d (\xi_m^d - \xi_n^d) \big) 
\end{equation}
and, for RoPEPool, we can rewrite (\ref{eq:ropeplus_attention_approx}) as:
\begin{equation} \label{eq:ropepool_toy_equation}
    a_{mn} \approx \sum_{d=1}^{D/2} 2 \cos \big( \psi_q^d + f_d \xi_m^d - \frac{ \pi}{4} \big) \cos \big( \psi_k^d + f_d \xi_n^d - \frac{\pi}{4} \big) 
\end{equation}
We give the derivations of (\ref{eq:rope_toy_equation}), (\ref{eq:fstripe1_toy_equation}) and (\ref{eq:ropepool_toy_equation}) in the supplementary materials.
We can comment on the expressivity of PE-enriched attention by studying the relationship between content ($\psi_q^d, \psi_k^d$) and context ($\xi_m^d, \xi_n^d$) in these three equations. 

RoPE (\ref{eq:rope_toy_equation}) leads to a linear interaction between content and context. 
F-StrIPE$_1$ (\ref{eq:fstripe1_toy_equation}) separates content and context information and models their interaction through an AND gate-like mechanism. A high score is only achieved when both the content and context of the query and the key match. 

What we can notice about (\ref{eq:rope_toy_equation}) and (\ref{eq:fstripe1_toy_equation}) is that both RoPE and F-StrIPE$_1$ model the contribution of positional information to the input in complementary ways. RoPE does this by using a phase-shift and F-StrIPE$_1$ does this by using a gain.
In contrast, RoPEPool mixes the input-based and position-based information for the query and key, instead of keeping them separate. Thus, this interaction depends not only on the difference in content and context, as was the case for RoPE (\ref{eq:rope_toy_equation}) and F-StrIPE$_1$ (\ref{eq:fstripe1_toy_equation}), but also on the identity of the content and context information. As hinted earlier, this introduces asymmetry into the picture, giving a richer interaction between content and context.

\begin{figure}[!b]
\centering
\includegraphics[width=2.5in]{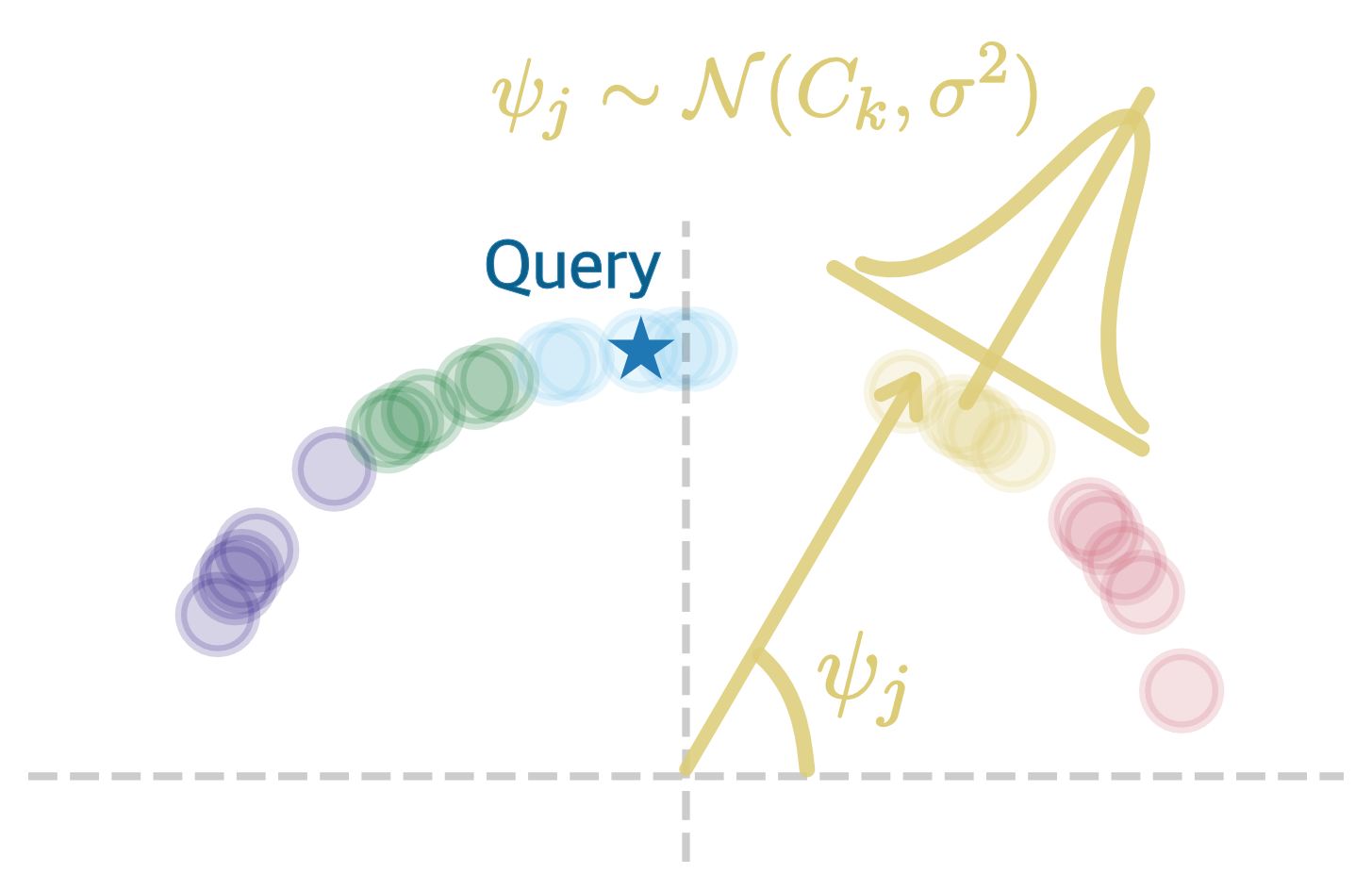}
\caption{Generating a toy dataset to study the characteristics of different positional encoding methods for the $D=2$ setting}
\label{fig:dataset}
\end{figure}

To further clarify these ideas, we simulate a simple toy example with a minimal setting of $D=2$ for (\ref{eq:rope_toy_equation}), (\ref{eq:fstripe1_toy_equation}) and (\ref{eq:ropepool_toy_equation}).
We devise a synthetic dataset by choosing $N$ context angles, $C_1, C_2, ..., C_N$, distributed evenly over the interval $(0, \pi)$. Each $C_i$ specifies the mean of a Gaussian distribution $\mathcal{D}_i$ that generates the samples of the $i^{th}$ context, as shown in Fig. \ref{fig:dataset}, where we set $N=5$. We create a dataset of $P$ points $\{\mathbf{x}_1, \mathbf{x}_2, ..., \mathbf{x}_P$\}, where each $\mathbf{x}_j$ is naturally associated with an angular coordinate $\psi_j$ that has been generated from one of the $N$ Gaussians as $\psi_j \sim \mathcal{N}(C_k, \sigma^2), k \in \{1, ..., N\}$. Further, $\mathbf{x}_j$ is also associated with $C_k$, the mean of the distribution it was generated from. Thus, for each point $\mathbf{x}_j$, $\psi_j$ defines its \textit{content}, capturing low-level, high-frequency information, and $\xi_j = C_k$ defines its \textit{context}, capturing high-level, low-frequency information.

\begin{figure}[!b]
\centering
\includegraphics[width=3.5in]{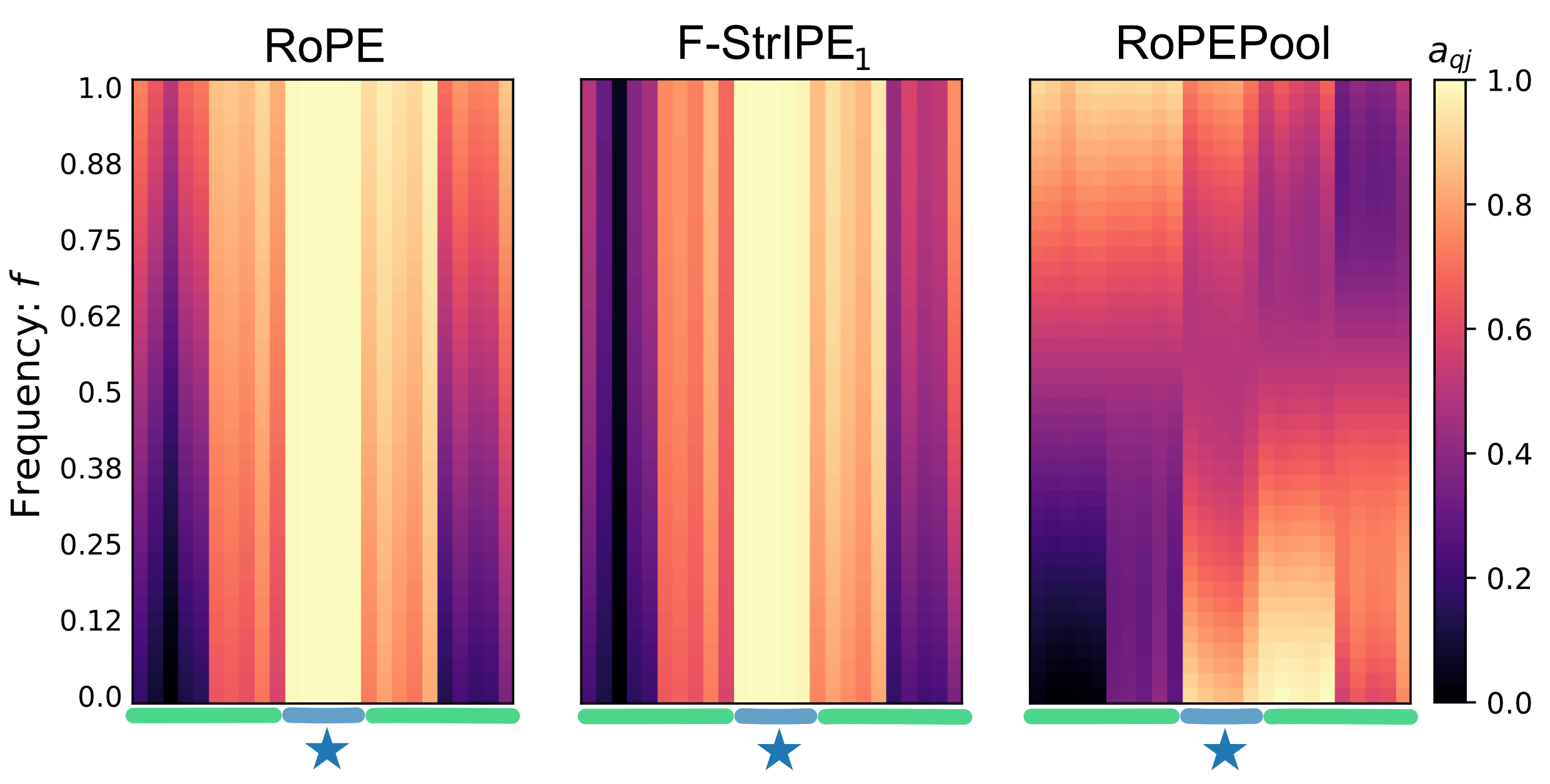}
\caption{Different PE methods exhibit different trade-offs between content and context}
\label{fig:results_synthetic}
\end{figure}

We designate one of the $P$ points as our `query' $\mathbf{x}_q$, which we mark with a star symbol in Fig. \ref{fig:dataset}. We are interested in measuring the score or similarity, given by $a_{qj}$, between $\mathbf{x}_q$ and all the points in our dataset, using (\ref{eq:rope_toy_equation}), (\ref{eq:fstripe1_toy_equation}) and (\ref{eq:ropepool_toy_equation}). We show the results in Fig. \ref{fig:results_synthetic}. The query $\mathbf{x}_q$ is marked with a star on the X-axis and we indicate which datapoints belong to the same context as $\mathbf{x}_q$ (marked in blue; we call these \textit{same-context keys} in our discussion below) and which datapoints belong to a different context (marked in green; we call these \textit{different-context keys}). 

We observe that F-StrIPE$_1$ exhibits attention values that are weakly modulated by $f$ and strongly modulated by the interaction between content and context. Discriminability between same-context keys and different-context keys with respect to the query remains robust for the full range of $f$ values. As discussed earlier, high values are achieved only when both the content and context of the key match those of the query, which can be seen clearly in Fig. \ref{fig:results_synthetic}.
For RoPE, this discriminability weakens as positional information starts to have an influence on attention, i.e., as $f$ increases.
Thus, even though RoPE and F-StrIPE$_1$ share the same expressivity, increasing positional information effectively \textit{blurs} the differences between the query and keys for RoPE, which is not the case for F-StrIPE$_1$.

RoPEPool, on the other hand, exhibits an entirely different kind of behaviour to F-StrIPE$_1$ and RoPE. 
For a fixed value of $f$, RoPEPool potentiates the scores of the keys in one $\psi_j$ range and depresses the scores of the keys under another $\psi_j$ range. As $f$ changes, these ranges also change. Thus, depending on the value of $f$, which we set to be learnable, RoPEPool can select a set of potentiated keys where the high scores concentrate and a set of depressed keys where the low scores concentrate. This is unlike RoPE and F-StrIPE$_1$, where the highest scores are always concentrated in the same-context keys. Furthermore, due to the left-right asymmetricity of the heatmap of RoPEPool, keys at a distance of $-\theta$ from the query get a different score than keys at a distance of $+\theta$. 
This is starkly different from RoPE and F-StrIPE$_1$, where both these sets of keys get the same score for all values of $f$. This allows us to model \textit{causality} with RoPEPool, which is a particularly salient feature in temporal sequences, such as musical signals.

Our analysis suggests that, in terms of expressivity, RoPE is the least expressive and RoPEPool is the most expressive, with F-StrIPE$_1$ occupying the middle rank. We hypothesize that RoPEPool should perform the best among these three methods on problems that require asymmetric, causal information.

\section{Experiments} \label{section:experiments}



\subsection{Dataset and Input Representation}

We use the Chinese POP909 dataset~\cite{wang_pop909_2020} for our experiments. Each MIDI file in this dataset contains 3 tracks: melody, bridge and piano. The bridge acts as a second melody track, supporting the development of the themes in the melody track. The piano track contains polyphonic accompaniment to the melody and bridge tracks. Each file is also paired with fine-grained multi-resolution structural annotations~\cite{dai_automatic_2020}. For this paper, following previous findings~\cite{agarwal_fast_2025}, we only use chord annotations for structural information, which uses a resolution of quarter-note. We use the POP909 alignment dataset~\cite{agarwal_structure_2024} to correctly sync the structural labels with the input.
We work with the MIDI data in the form of binary pianorolls $\mathbf{X} \in \mathbb{B}^{(n_\text{tracks} \times 128) \times n_\text{time}}, \mathbb{B} = \{0, 1\}$, where $n_\text{tracks}$ is the number of tracks and $n_\text{time}$ is the number of timesteps in the pianoroll. The POP909 dataset has approximately 400 unique chords. For each chord, we have access to two types of information: the scalar root note and the 12-dimensional chroma vector, which indicates the pitches that make up the chord. We investigate both token-based and vector-based representations of this information, as we will describe in later sections.

\subsection{Task: Melody Harmonization}

As input, the model is given the melody and bridge tracks $\big[ \mathbf{x}_n \in \mathbb{B}^{(n_\text{tracks} - 1) \times 128} \big]$, with $n \in \{ 1, ..., n_\text{time}\}$. As output, the model produces all three tracks $\big[ \mathbf{y}_n \in \mathbb{B}^{n_\text{tracks} \times 128} \big]$. Note that, here, we are in an extreme parallel decoding regime, where we require our model to produce the full accompaniment track at once and do not allow our model to autoregressively condition later predictions on earlier predictions.
We train our models on sequences of length 16 bars. We test two settings: the \textit{in-domain} setting, with sequences of length 16 bars, and the \textit{length generalization} setting, with sequences of length 64 bars.

\subsection{Model and Training}
We use a 2-layer causal encoder Transformer with 4 heads and 512 model dimension. We train for 15 epochs with a batch size of 8, gradient clipping and curriculum learning~\cite{bengio_curriculum_2009}. We use two learning rate schedulers: a linear warmup, applied for the first three epochs, and a decay, applied for each epoch.
We do a grid-search for two hyperparameters: learning rate (choices: $\{ 1, 5, 10 \} \times 0.0001$) and post-processing binarization strategy~\cite{agarwal_structure_2024} (choices: thresholding, thresholding with merge). The first binarization strategy uses a fixed threshold on the logits to obtain binarized outputs. The second binarization strategy, first, uses a fixed threshold to binarize the logits and, second, allows the gap between two binarized notes to be filled when this gap is less than a minimum distance.

\subsection{Types of Contextual Information} \label{ssection:context_type}

To understand how context influences model performance, we represent positional information in several ways, varying how much local versus global information the context contains.

To encode fully \textit{local} information, we use \textit{time} (TIME). Since each time-step is assigned a unique token, this representation assumes all inputs in a sample are independent.

To encode fully \textit{global} information, we use \textit{binary chroma chord vectors} (BIN), which contain ones for the notes that occur in a chord and zeros for those that don't. This accounts for the harmonic content of each chord, allowing us to identify repetitions of harmonically-similar chords across different samples at the level of the dataset.

Between these two extremities, we use two \textit{blended} chord representations, which meld together local and global information. The first is \textit{repetition-based chord tokens} (REP). Here, we first represent each chord as a unique token, taking into account both its root note and chroma vector. Then, to reduce the number of tokens and focus on repetitions within samples, we re-number the tokens in each sample based on their numerical order. The second is \textit{key-matched chord tokens} (KEY). Here, we use the fact that the POP909 dataset provides the key of each song. Using this, we `normalize' each chord with respect to its song's key, mapping the chord to its counterpart in either the C:major scale or the C:minor scale. For example, the E:minor chord in D:major scale is mapped to the D:minor chord in C:major scale. Thus, we focus on the function of each chord across samples, while blurring harmonic correlations between chords and a song's pitches.

It is worth noting that we selected three hand-designed representations, based in concepts from music theory, to simplify our experiments, but optimization-based representations~\cite{wang_supervised_2022} can also be used with the PE methods studied in this paper.

\subsection{Baselines} \label{ssection:baselines}


To construct baselines for each type of contextual information described above, we investigate three variants of RoPE. The first variant, RoPE(A), is the standard method from literature~\cite{su_roformer_2024}, where the frequencies are fixed, following an exponential distribution, and all heads contain the same frequencies. The second variant, RoPE(B), modifies RoPE(A) by allowing different heads to have different frequencies. The third variant, RoPE(C), builds on RoPE(B) by allowing frequencies to be learned as parameters.
For each contextual information type, we choose the RoPE variant that performs the best.
Thereafter, we compare these chosen RoPE baselines with F-StrIPE$_1$ and RoPEPool for each context type. For F-StrIPE$_1$ and RoPEPool, we use a system analogous to RoPE(C), where heads have different, learnable frequencies.

\subsection{Evaluation}

Following existing work~\cite{agarwal_structure_2024,agarwal_fast_2025}, we choose four objective metrics from the literature to compare the ground truth and the generated music. Each focuses on one axis of musical quality.

We use the \textit{Self-Similarity Matrix Distance (SSMD)}~\cite{wu_musemorphose_2021} to evaluate large-scale and small-scale structural properties. We compute chroma onset vectors, giving the number of onsets per chroma per half-measure, for the target and the prediction. For each chroma onset vector, we calculate its self-similarity matrix (SSM) by taking pairwise cosine similarities between all elements. We take the absolute difference between the target's SSM and the prediction's SSM, giving us a matrix with entries varying between 0 and 2. The SSMD is the mean of this matrix, linearly scaled to vary between 0 and 100.

We use \textit{Chroma Similarity (CS)}~\cite{wu_musemorphose_2021} to measure melodic consistency. The harmonic match between each half-measure in the target and its counterpart in the prediction is the cosine similarity between their corresponding chroma onset vectors, derived as described above. The mean of these cosine similarities, varying between -1 and 1, is linearly scaled to vary between -100 and 100, and this gives us the CS.

We use \textit{Grooving pattern Similarity (GS)}~\cite{wu_jazz_2020} to assess rhythmic consistency. The grooving pattern of a pianoroll is a vector that encodes a 1 for the quarter-notes where onsets occur and 0 for those where no onsets occur. The XOR function is applied to the grooving patterns of the target and the prediction, after which we take the mean of the resulting vector, giving us a number between 0 and 1. We linearly scale this to vary from 0 to 100, giving us the GS.

To estimate polyphonicity, we use \textit{Note Density Distance (NDD)}~\cite{agarwal_structure_2024, haki_real_2022}. For each $16^{\text{th}}$-note of the target and prediction, we calculate the total number of pitches that occur. The NDD is the average percentage of missing pitches in the prediction, with the number of pitches in the target giving us the maximum possible value. It varies between 0 and 100.

Note that all four metrics are scale-invariant to sequence length, making the results from the in-domain and length generalization scenarios comparable with each other.

\section{Results} \label{section:results}

\begin{table}[!b]
    \centering
    \begin{tabular}{cc}
        \hline
        Context type & Mutual information with data distribution (in nats) \\ \hline \hline
        TIME & 0.0085 \\
        REP & 0.0436 \\
        KEY & 0.0476 \\
        BIN & 1.0075 \\ \hline
    \end{tabular}
    \caption{Mutual information between content and context}
    \label{tab:mi}
\end{table}

In Table \ref{tab:pe_variants}, we first present the results on the three variants of RoPE described in Section \ref{ssection:baselines}. From these, we select the RoPE baseline with the best performance for each context type and compare this with F-StrIPE$_1$ and RoPEPool. These results are shown in Table \ref{tab:results}. 

For each PE method in both tables, we train models using five seeds and report the mean and standard deviation for each metric.
For Table \ref{tab:results}, we underline the PE method with the best mean score for each context type. However, we check if there is a statistically significant difference between the underlined method and other methods within the same context by using two-sided t-tests~\cite{moore_stats_1999}. For a pair of PE methods, we first use a Levene test~\cite{levene_robust_1961} to check whether the pair of score distributions have the same variance. If this is the case, we use Student's t-test to make the comparison, and, if not, we use Welch's t-test~\cite{welch_generalization_1947}. We use a $\ast$ with the underlined score to indicate that there is a significant difference ($p < 0.05$) between the best (underlined) mean score and the next-best mean score. If we do not find any significant difference in this comparison, we use a $\dagger$ to indicate that there is a statistically significant difference between the best mean score and the third-best mean score. This is to gain a fine-grained understanding on how different PE methods perform relative to each other within a given context.
In Table \ref{tab:results}, we additionally also highlight the best PE method among the different underlined PE methods in bold in order to indicate which combination of context and PE is optimal.


\subsection{Ordering Context Type by Information Content} \label{ssection:context_mi}

Before we describe the results in Tables \ref{tab:pe_variants} and \ref{tab:results}, let us formalize the notion of `local versus global' with regards to context type, which we introduced in Section \ref{ssection:context_type}. To do this, we calculate the mutual information (or MI score) between the distribution of positional information and the data distribution. We convert the binary pianorolls into an array of individual pitches and match each pitch with its corresponding positional index, which changes depending on the context type and the timestep in which the pitch occurs in the pianoroll. Then, we calculate the MI score between the input, represented as pitches, and their positions. We report this score for each context type in Table \ref{tab:mi}. Intuitively, the mutual information between these two distributions measures how well we can predict one from the other. We expect that a higher MI score makes it easier for the model to learn the data distribution by leveraging contextual information via positional encoding.
\begin{table*}[t]
\centering
\resizebox{2.05\columnwidth}{!}{%
\begin{tabular}{ccccccccc}
\hline
\multicolumn{1}{c|}{\multirow{3}{*}{\textbf{Configuration}}} & \multicolumn{4}{c|}{\textbf{Train = 16 bars; Test = 16 bars}}   & \multicolumn{4}{c}{\textbf{Train = 16 bars; Test = 64 bars}}        \\ 
\multicolumn{1}{c|}{}        & \textbf{CS}  & \textbf{SSMD}  & \textbf{GS}    & \multicolumn{1}{c|}{\textbf{NDD}}  & \textbf{CS}    & \textbf{SSMD}  & \textbf{GS}  & \textbf{NDD}   \\
\multicolumn{1}{c|}{}                               & $\uparrow$   & $\downarrow$ & $\uparrow$   & \multicolumn{1}{c|}{$\downarrow$} & $\uparrow$   & $\downarrow$ & $\uparrow$    & $\downarrow$ \\ \hline \hline
\multicolumn{9}{c}{\textbf{Time}}                                                                                                                                                                          \\ \hline
\multicolumn{1}{c|}{\textbf{RoPE(A)}}                        & 2.44 ± 0.06  & \underline{14.65 ± 0.0} & 7.46 ± 0.05  & \multicolumn{1}{c|}{94.1 ± 0.04}  & 5.83 ± 3.27  & 13.8 ± 0.0  & 18.27 ± 9.1   & 91.79 ± 0.84 \\
\multicolumn{1}{c|}{\textbf{RoPE(B)}}                        & 1.61 ± 0.74  & 14.66 ± 0.01 & 6.72 ± 0.71  & \multicolumn{1}{c|}{94.38 ± 0.27} & 2.23 ± 0.09  & 13.8 ± 0.0   & 8.44 ± 0.21   & 92.72 ± 0.03 \\
\multicolumn{1}{c|}{\textbf{RoPE(C)}}                        & \underline{4.5 ± 4.16}   & \underline{14.65 ± 0.02} & \underline{10.42 ± 5.32} & \multicolumn{1}{c|}{\underline{93.82 ± 1.01}} & \underline{9.1 ± 0.49}$^\dagger$   & \underline{13.79 ± 0.0}$^\ast$ & \underline{26.78 ± 1.19}$^\dagger$  & \underline{91.26 ± 0.2}$^\dagger$  \\ \hline
\multicolumn{9}{c}{\textbf{Repetition-based chord tokens}}                                                                                                                                                 \\ \hline
\multicolumn{1}{c|}{\textbf{RoPE(A)}}                        & 5.42 ± 5.51  & 14.61 ± 0.06 & 10.89 ± 6.65 & \multicolumn{1}{c|}{92.75 ± 2.27} & 15.42 ± 1.18 & 13.67 ± 0.01 & 42.4 ± 3.23   & 86.77 ± 0.92 \\
\multicolumn{1}{c|}{\textbf{RoPE(B)}}                        & \underline{18.37 ± 1.11}$^\ast$ & \underline{14.36 ± 0.03}$^\ast$ & \underline{25.63 ± 1.76}$^\ast$ & \multicolumn{1}{c|}{\underline{86.13 ± 0.57}$^\ast$} & \underline{18.17 ± 1.92}$^\ast$ & \underline{13.46 ± 0.02}$^\ast$ & \underline{49.56 ± 3.12}$^\ast$  & \underline{82.9 ± 0.54}$^\ast$  \\
\multicolumn{1}{c|}{\textbf{RoPE(C)}}                        & 4.59 ± 5.16  & 14.59 ± 0.11 & 9.76 ± 5.6   & \multicolumn{1}{c|}{92.84 ± 2.57} & 9.46 ± 7.28  & 13.62 ± 0.15 & 27.27 ± 20.17 & 88.35 ± 4.06 \\ \hline \hline
\multicolumn{9}{c}{\textbf{Key-matched chord tokens}}                                                                                                                                                      \\ \hline
\multicolumn{1}{c|}{\textbf{RoPE(A)}}                        & 11.05 ± 5.62 & \underline{14.49 ± 0.09} & 16.89 ± 6.59 & \multicolumn{1}{c|}{\underline{89.64 ± 2.56}} & 14.59 ± 5.24 & 13.63 ± 0.08 & 39.02 ± 12.34 & 86.82 ± 2.35 \\
\multicolumn{1}{c|}{\textbf{RoPE(B)}}                        & 11.85 ± 4.6  & 14.5 ± 0.07  & 18.49 ± 5.52 & \multicolumn{1}{c|}{89.85 ± 1.99} & \underline{16.2 ± 2.21}  & 13.63 ± 0.01 & \underline{43.67 ± 3.95}  & \underline{85.87 ± 0.39} \\
\multicolumn{1}{c|}{\textbf{RoPE(C)}}                        & \underline{12.03 ± 4.51} & \underline{14.49 ± 0.07} & \underline{18.57 ± 5.37} & \multicolumn{1}{c|}{89.87 ± 1.93} & 15.84 ± 1.09 & \underline{13.62 ± 0.02} & 42.04 ± 2.65  & 86.68 ± 0.85 \\ \hline \hline
\multicolumn{9}{c}{\textbf{Binary chroma vectors}}                                                                                                                                                         \\ \hline
\multicolumn{1}{c|}{\textbf{RoPE(A)}}                        & 1.81 ± 0.34  & 14.64 ± 0.01 & 6.76 ± 0.35  & \multicolumn{1}{c|}{94.3 ± 0.14}  & 10.04 ± 4.21 & 13.74 ± 0.02 & 30.39 ± 11.72 & 89.44 ± 1.69 \\
\multicolumn{1}{c|}{\textbf{RoPE(B)}}                        & 1.9 ± 0.18   & 14.66 ± 0.01 & 7.07 ± 0.18  & \multicolumn{1}{c|}{94.29 ± 0.07} & 3.51 ± 2.56  & 13.8 ± 0.0   & 11.93 ± 7.12  & 92.31 ± 0.73 \\
\multicolumn{1}{c|}{\textbf{RoPE(C)}}                        & \underline{22.14 ± 0.56}$^\ast$ & \underline{13.98 ± 0.01}$^\ast$ & \underline{26.45 ± 0.52}$^\ast$ & \multicolumn{1}{c|}{\underline{79.25 ± 0.29}$^\ast$} & \underline{26.27 ± 0.44}$^\ast$ & \underline{13.13 ± 0.01}$^\ast$ & \underline{66.81 ± 0.6}$^\ast$   & \underline{75.02 ± 0.34}$^\ast$ \\ \hline
\end{tabular}
}
\caption{Comparing different RoPE baselines (Section \ref{ssection:baselines}) using different contexts (Section \ref{ssection:context_type})}
\label{tab:pe_variants}
\end{table*}
\begin{table*}[t]
\centering
\resizebox{2.05\columnwidth}{!}{%
\begin{tabular}{ccccccccc}
\hline
\multicolumn{1}{c|}{\multirow{3}{*}{\textbf{Configuration}}} & \multicolumn{4}{c|}{\textbf{Train = 16 bars; Test = 16 bars}}   & \multicolumn{4}{c}{\textbf{Train = 16 bars; Test = 64 bars}}        \\ 
\multicolumn{1}{c|}{}        & \textbf{CS}  & \textbf{SSMD}  & \textbf{GS}    & \multicolumn{1}{c|}{\textbf{NDD}}  & \textbf{CS}    & \textbf{SSMD}  & \textbf{GS}  & \textbf{NDD}   \\
\multicolumn{1}{c|}{}                               & $\uparrow$   & $\downarrow$ & $\uparrow$   & \multicolumn{1}{c|}{$\downarrow$} & $\uparrow$   & $\downarrow$ & $\uparrow$    & $\downarrow$ \\ \hline \hline
\multicolumn{9}{c}{\textbf{Time}}                                                                                                                                                                          \\ \hline
\multicolumn{1}{c|}{\textbf{RoPE(C)}}                        & 4.5 ± 4.16   & 14.65 ± 0.02  & \underline{10.42 ± 5.32} & \multicolumn{1}{c|}{93.82 ± 1.01} & \underline{9.1 ± 0.49}$^\ast$    & 13.79 ± 0.0   & \underline{26.78 ± 1.19}$^\ast$ & 91.26 ± 0.2  \\
\multicolumn{1}{c|}{\textbf{F-StrIPE$_1$}}                   & \underline{6.01 ± 2.67}  & \underline{14.64 ± 0.02}  & 11.9 ± 3.24  & \multicolumn{1}{c|}{\underline{93.2 ± 0.76}}  & 6.68 ± 1.12   & 13.79 ± 0.0   & 20.67 ± 2.59 & 91.4 ± 0.21  \\
\multicolumn{1}{c|}{\textbf{RoPEPool}}                       & 2.49 ± 2.64  & 14.66 ± 0.01  & 7.79 ± 3.32  & \multicolumn{1}{c|}{94.28 ± 0.68} & 7.23 ± 0.62   & \underline{13.79 ± 0.0}$^\ast$   & 22.13 ± 1.65 & \underline{91.18 ± 0.2}  \\ \hline \hline
\multicolumn{9}{c}{\textbf{Repetition-based chord tokens}}                                                                                                                                                 \\ \hline
\multicolumn{1}{c|}{\textbf{RoPE(B)}}                        & \underline{18.37 ± 1.11}$^\dagger$ & \underline{14.36 ± 0.03}$^\dagger$  & \underline{25.63 ± 1.76}$^\dagger$ & \multicolumn{1}{c|}{\underline{86.13 ± 0.57}$^\dagger$} & \underline{18.17 ± 1.92}$^\ast$  & \underline{13.46 ± 0.02}$^\ast$   & \underline{49.56 ± 3.12}$^\ast$ & \underline{82.9 ± 0.54}$^\ast$  \\
\multicolumn{1}{c|}{\textbf{F-StrIPE$_1$}}                   & 17.71 ± 0.97 & 14.37 ± 0.03  & 24.82 ± 1.22 & \multicolumn{1}{c|}{86.81 ± 0.71} & 13.85 ± 1.18  & 13.59 ± 0.03   & 37.66 ± 1.93 & 86.47 ± 0.66 \\
\multicolumn{1}{c|}{\textbf{RoPEPool}}                       & 14.23 ± 1.13 & 14.47 ± 0.02  & 20.87 ± 1.86 & \multicolumn{1}{c|}{88.81 ± 0.29} & 11.67 ± 0.58  & 13.7 ± 0.01   & 34.11 ± 1.12 & 88.04 ± 0.52 \\ \hline \hline
\multicolumn{9}{c}{\textbf{Key-matched chord tokens}}                                                                                                                                                      \\ \hline
\multicolumn{1}{c|}{\textbf{RoPE(C)}}                        & 12.03 ± 4.51 & 14.49 ± 0.07  & 18.57 ± 5.37 & \multicolumn{1}{c|}{89.87 ± 1.93} & 15.84 ± 1.09  & 13.62 ± 0.02   & 42.04 ± 2.65 & 86.68 ± 0.85 \\
\multicolumn{1}{c|}{\textbf{F-StrIPE$_1$}}                   & 19.73 ± 0.4  & \underline{14.1 ± 0.02}$^\ast$  & 24.08 ± 0.48 & \multicolumn{1}{c|}{\underline{81.41 ± 0.27}$^\dagger$} & \underline{23.6 ± 0.51}$^\dagger$   & \underline{13.2 ± 0.05}$^\dagger$    & \underline{60.87 ± 1.28}$^\dagger$ & \underline{77.47 ± 0.35}$^\dagger$ \\
\multicolumn{1}{c|}{\textbf{RoPEPool}}                       & \underline{21.05 ± 1.22}$^\dagger$ & 14.18 ± 0.06  & \underline{26.05 ± 1.53}$^\ast$ & \multicolumn{1}{c|}{82.46 ± 0.96} & 23.11 ± 1.43  & 13.28 ± 0.09   & 59.74 ± 2.46 & 78.69 ± 1.37 \\ \hline \hline
\multicolumn{9}{c}{\textbf{Binary chroma vectors}}                                                                                                                                                         \\ \hline
\multicolumn{1}{c|}{\textbf{RoPE(C)}}                        & 22.14 ± 0.56 & 13.98 ± 0.01  & 26.45 ± 0.52 & \multicolumn{1}{c|}{79.25 ± 0.29} & 26.27 ± 0.44  & 13.13 ± 0.01   & 66.81 ± 0.6  & 75.02 ± 0.34 \\
\multicolumn{1}{c|}{\textbf{F-StrIPE$_1$}}                   & 24.91 ± 0.35 & \underline{\textbf{13.73 ± 0.03}}$^\dagger$  & 28.19 ± 0.31 & \multicolumn{1}{c|}{77.15 ± 0.33} & 28.3 ± 0.81   & \underline{\textbf{12.95 ± 0.04}}$^\dagger$    & 71.35 ± 0.99 & 73.97 ± 0.48 \\
\multicolumn{1}{c|}{\textbf{RoPEPool}}                       & \underline{\textbf{25.83 ± 1.02}}$^\dagger$ & 13.74 ± 0.07  & \underline{\textbf{29.11 ± 0.97}}$^\dagger$ & \multicolumn{1}{c|}{\underline{\textbf{76.73 ± 1.09}}$^\dagger$} & \underline{\textbf{28.6 ± 0.8}}$^\dagger$    & 12.99 ± 0.06   & \underline{\textbf{72.56 ± 1.54}}$^\dagger$ & \underline{\textbf{73.77 ± 1.23}} \\ \hline \hline
\end{tabular}
}
\caption{Comparing different PE methods using different contexts. We chose the best RoPE baseline from Table \ref{tab:pe_variants} to compare against F-StrIPE$_1$ and RoPEPool}
\label{tab:results}
\end{table*}

\subsection{Comparing PE Methods within Context Types} \label{ssection:pe_context_description}

For each context type, we first describe the results for selecting a RoPE baseline. Then, we compare the results of the chosen RoPE baseline with those of the other PE methods.

For Table \ref{tab:pe_variants}, in the in-domain setting, we see statistically significant results for two context types, namely, REP and BIN. In the length generalization setting, we see this for three context types, namely TIME, REP and BIN. For REP, RoPE(B) performs the best and, for BIN, RoPE(C) performs the best, in both settings. For TIME, RoPE(C) performs the best, particularly in the length generalization setting. Finally, for KEY, there is no statistically significant difference between the different PE methods, with RoPE(B) or RoPE(C) achieving the best mean score in most metrics for both settings.



Similar to Table \ref{tab:pe_variants}, TIME in Table \ref{tab:results} does not have statistically significant results in the in-domain scenario, but has statistically significant results in the length generalization scenario. Similarly, for KEY, F-StrIPE$_1$ and RoPEPool jointly perform best, although, unlike Table \ref{tab:pe_variants}, the difference with RoPE is statistically significant. In REP, RoPE(B) continues its win from Table \ref{tab:pe_variants} and outperforms the other PE methods. For BIN, RoPEPool performs the best, with F-StrIPE$_1$ showing similar performance with a slightly worse mean score.

\subsection{Utility of Structural Information} \label{ssection:utility_structure}

In Table \ref{tab:results}, on comparing the best PE methods from each context, we find that BIN, encoding global musically-relevant information, performs the best, and TIME performs the worst, with the blended representations, namely, REP and KEY, occupying the middle ranks. In both the in-domain and length generalization scenarios, KEY enables higher performance than REP, although the difference between the two is not significant for GS in the in-domain setting ($p > 0.5$).

\subsection{Length Generalization} \label{ssection:length_gen_scenario}

In Table \ref{tab:results}, with the exception of RoPE applied to REP, all methods and context types lead to an amelioration of scores from the in-domain scenario to the length generalization scenario, which is particularly dramatic for GS in BIN. Testing on out-of-domain samples tends to augment the score differences between various PE methods, especially for GS.

\section{Discussion} \label{section:discussion}

\subsection{Explaining the Success of Structural Positional Encoding with Mutual Information} \label{ssection:success_mi}

Table \ref{tab:mi} formalizes our intuition that TIME contains no information about connections within a sample, since it assumes that all time-steps are contextually independent, while REP, KEY and BIN introduce progressively stronger contextual information about the input distribution. Note that even though KEY has a higher MI score than REP, they both fall within the same order of magnitude. We contrast this with the comparison between TIME, REP/KEY and BIN, which lead to MI scores at increasing orders of magnitude. This captures the intuition that moving from local to blended to global information in context leads to fundamental changes in what additional insight can be gained with respect to the input distribution, whereas moving within the same informational regime (e.g., REP to KEY) introduces subtler interactions.

By combining our observations from Sections \ref{ssection:context_mi} and \ref{ssection:utility_structure}, we argue that this is precisely what makes task-specific priors, such as structural information, better than generic priors, such as time information, in positional encoding. This point is further strengthened by the fact that the ranking in MI scores (Table \ref{tab:mi}) matches precisely the ranking in performance with respect to context (Table \ref{tab:results}). These findings could also explain how and why harnessing information about the structural properties of data has been found to aid learnability and improve performance in other generative architectures such as diffusion models, both with synthetic~\cite{favero_compositional_2025} and real data~\cite{mei_power_2025}.

Having translated our intuitions into a quantitative form, we can now analyze the empirical results from Tables \ref{tab:pe_variants} and \ref{tab:results} in further detail while bearing these ideas in mind.

\subsection{A Phenomenological Perspective on Performance vis-\'{a}-vis Information Content}

\begin{figure}[!b]
\centering
\includegraphics[width=2.5in]{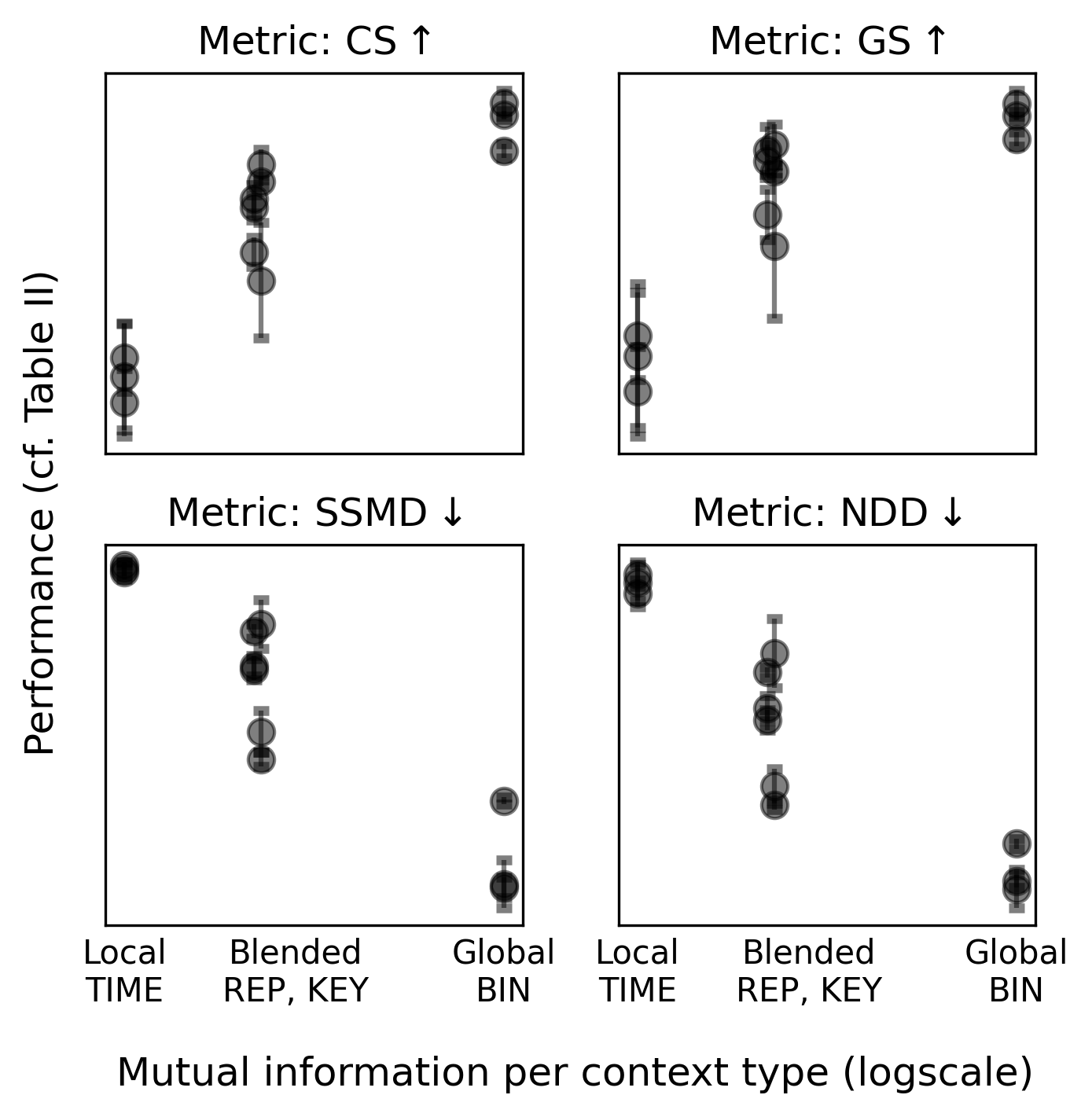}
\caption{Performance as a function of how much information about the data distribution is contained in the positional information; drawn from the in-domain results in Table \ref{tab:results}}
\label{fig:mi}
\end{figure}

As presented in Section \ref{ssection:pe_context_description}, the impact of context on performance has two facets: the relative performance of different PE methods (i.e., which method performs best) and the difference between these methods (i.e., is the best performance significantly different from the others). To disentangle them, we can plot the in-domain performance of each PE method on each metric from Table \ref{tab:results} against the MI scores from Table \ref{tab:mi}. The results are displayed in Fig. \ref{fig:mi}.

First, this plot, along with Table \ref{tab:results}, suggests the existence of a transition from no differentiation between PE methods, in the no-information setting (TIME), to simpler PE methods performing better than more complex PE methods, in limited-information scenarios (REP and KEY), to more complex PE methods performing better than simpler PE methods, in high-information scenarios (BIN).
Thus, as contextual information in PE increases, there is also the \textit{emergence of differentiation} between PE methods.
We can see this in Fig. \ref{fig:mi} by comparing TIME, where the different PE methods largely overlap, with other context types, where the PE methods develop distinct distributions.
Second, it shows that the mutual information between data and its positional representation correlates strongly with performance, adding further evidence to our analysis in Section \ref{ssection:success_mi}. The more information that is encoded about the input distribution with positional encoding, the better is the performance of the model. Finally, it shows that imperfect mutual information (REP/KEY) leads to a large spread of results. This region of positional information is defined by large error bars and a wide range of mean scores.

These observations enable us to characterize the interaction between context type and comparative performance of PEs using two interrelated phenomena. First, increasing the relevance of positional information with respect to the input diversifies performance, such that differences between PEs become more evident as the context distribution starts to encode more information about the content distribution. This indicates that empirical works comparing different PE methods might see none or highly dampened differences in the performance of different PEs, not due to the PE's inherent properties, but because the PE does not encode enough information, for example, in the typical case where TIME is used.
Second, there exists an intermediary stage of relevance, with imperfect mutual information between the positions and the input, that exhibits a large variance in performance among the different PE methods. This indicates that, under imperfect structural information, performance is highly initialization-dependent and optimization in such cases might be unstable. This might also explain why, for example, on REP, RoPE(B) does better than RoPE(C) in Table \ref{tab:pe_variants} and better than F-StrIPE$_1$ and RoPEPool in Table \ref{tab:results}.

\section{Conclusion}

In this paper, we compared two families of efficient positional encoding methods: those based on Random Fourier Features and those based on rotation matrices. We began by unifying these families under a common framework of PE-enriched attention, which we also used to develop a novel PE called RoPEPool. With a toy example, we studied the expressiveness of RoPE, F-StrIPE$_1$ and RoPEPool and showed that, unlike RoPE and F-StrIPE$_1$, RoPEPool can extract causal relationships from temporal sequences.
To assess the validity of our ideas, we used melody harmonization, a symbolic music generation task.
We also varied the type and representation of positional information used, ordering them by how much they tell us about the structure of the data. We found that combining RoPEPool with highly-informative structural information performs the best, matching our conclusions from the toy example.
We went beyond these task-specific findings to clarify the role of information about data structure in PE. We used the mutual information between the content distribution (via the input) and the context distribution (via the PE) to explain why rich contexts aid learnability. We provided a phenomenological account of how the performance of different PEs varies as context information increases, enriching our comparative study. 

\section*{Acknowledgements}

We thank Emmanouil Benetos for helpful discussions around variations of chord representations.


\bibliographystyle{IEEEtran}
\bibliography{refs}

\begin{thebibliography}{10}
\providecommand{\url}[1]{#1}
\csname url@samestyle\endcsname
\providecommand{\newblock}{\relax}
\providecommand{\bibinfo}[2]{#2}
\providecommand{\BIBentrySTDinterwordspacing}{\spaceskip=0pt\relax}
\providecommand{\BIBentryALTinterwordstretchfactor}{4}
\providecommand{\BIBentryALTinterwordspacing}{\spaceskip=\fontdimen2\font plus
\BIBentryALTinterwordstretchfactor\fontdimen3\font minus
  \fontdimen4\font\relax}
\providecommand{\BIBforeignlanguage}[2]{{%
\expandafter\ifx\csname l@#1\endcsname\relax
\typeout{** WARNING: IEEEtran.bst: No hyphenation pattern has been}%
\typeout{** loaded for the language `#1'. Using the pattern for}%
\typeout{** the default language instead.}%
\else
\language=\csname l@#1\endcsname
\fi
#2}}
\providecommand{\BIBdecl}{\relax}
\BIBdecl

\bibitem{kaplan_scaling_2020}
\BIBentryALTinterwordspacing
J.~Kaplan, S.~McCandlish, T.~Henighan, T.~B. Brown, B.~Chess, R.~Child,
  S.~Gray, A.~Radford, J.~Wu, and D.~Amodei, ``Scaling laws for neural language
  models,'' \emph{arXiv}, 2020. [Online]. Available:
  \url{https://arxiv.org/abs/2001.08361}
\BIBentrySTDinterwordspacing

\bibitem{wu_jazz_2020}
S.~Wu and Y.~Yang, ``The {Jazz} {Transformer} on the {Front} {Line}:
  {Exploring} the {Shortcomings} of {AI}-composed {Music} through
  {Quantitative} {Measures},'' \emph{Conference of the International Society
  for Music Information Retrieval (ISMIR)}, 2020.

\bibitem{ji_survey_2023}
S.~Ji, X.~Yang, and J.~Luo, ``A survey on deep learning for symbolic music
  generation: Representations, algorithms, evaluations, and challenges,''
  \emph{ACM Computing Surveys}, vol.~56, no.~1, 2023.

\bibitem{richard_model_2024}
G.~Richard, V.~Lostanlen, Y.-H. Yang, and M.~Müller, ``Model-based deep
  learning for music information research: Leveraging diverse knowledge sources
  to enhance explainability, controllability, and resource efficiency,''
  \emph{IEEE Signal Processing Magazine}, vol.~41, no.~6, pp. 51--59, 2024.

\bibitem{bhandari_motifs_2024}
K.~Bhandari and S.~Colton, ``Motifs, {P}hrases, and {B}eyond: {T}he {M}odelling
  of {S}tructure in {S}ymbolic {M}usic {G}eneration,'' \emph{International
  Conference on Computational Intelligence in Music, Sound, Art and Design},
  2024.

\bibitem{agarwal_structure_2024}
M.~Agarwal, C.~Wang, and G.~Richard, ``Structure-{I}nformed {P}ositional
  {E}ncoding for {M}usic {G}eneration,'' \emph{International Conference on
  Acoustics, Speech and Signal Processing (ICASSP)}, 2024.

\bibitem{yi_popmag_2020}
Y.~Ren, J.~He, X.~Tan, T.~Qin, Z.~Zhao, and T.-Y. Liu, ``{PopMAG}: {Pop}
  {Music} {Accompaniment} {Generation},'' \emph{ACM International Conference on
  Multimedia (MM)}, 2020.

\bibitem{guo_domain_2023}
Z.~Guo, J.~Kang, and D.~Herremans, ``{A Domain-Knowledge-Inspired Music
  Embedding Space and a Novel Attention Mechanism for Symbolic Music
  Modeling},'' \emph{AAAI Conference on Artificial Intelligence (AAAI)}, 2023.

\bibitem{liu2022symphony}
J.~Liu, Y.~Dong, Z.~Cheng, X.~Zhang, X.~Li, F.~Yu, and M.~Sun, ``{Symphony
  Generation with Permutation Invariant Language Model},'' \emph{Conference of
  the International Society for Music Information Retrieval (ISMIR)}, 2022.

\bibitem{tay_efficient_2022}
Y.~Tay, M.~Dehghani, D.~Bahri, and D.~Metzler, ``Efficient transformers: A
  survey,'' \emph{ACM Computing Surveys}, vol.~55, no.~6, 2022.

\bibitem{tsai_transformer_2019}
Y.-H.~H. Tsai, S.~Bai, M.~Yamada, L.-P. Morency, and R.~Salakhutdinov,
  ``{Transformer {D}issection: {A}n {U}nified {U}nderstanding for
  {T}ransformer{'}s {A}ttention via the {L}ens of {K}ernel},'' \emph{Empirical
  Methods in Natural Language Processing (EMNLP)}, 2019.

\bibitem{agarwal_fast_2025}
M.~Agarwal, C.~Wang, and G.~Richard, ``F-stripe: Fast structure-informed
  positional encoding for symbolic music generation,'' \emph{International
  Conference on Acoustics, Speech and Signal Processing (ICASSP)}, 2025.

\bibitem{liutkus_relative_2021}
A.~Liutkus, O.~C\'{\i}fka, S.-L. Wu, U.~Simsekli, Y.-H. Yang, and G.~Richard,
  ``Relative {P}ositional {E}ncoding for {T}ransformers with {L}inear
  {C}omplexity,'' \emph{International Conference on Machine Learning (ICML)},
  2021.

\bibitem{rahimi_random_2007}
A.~Rahimi and B.~Recht, ``Random features for large-scale kernel machines,''
  \emph{Annual Conference on Neural Information Processing Systems (NeurIPS)},
  2007.

\bibitem{su_roformer_2024}
J.~Su, M.~Ahmed, Y.~Lu, S.~Pan, W.~Bo, and Y.~Liu, ``Roformer: Enhanced
  transformer with rotary position embedding,'' \emph{Neurocomputing}, vol.
  568, p. 127063, 2024.

\bibitem{ke_rethinking_2021}
G.~Ke, D.~He, and T.~Liu, ``Rethinking positional encoding in language
  pre-training,'' \emph{International Conference on Learning Representations
  (ICLR)}, 2021.

\bibitem{chang_convolutions_2021}
T.~Chang, Y.~Xu, W.~Xu, and Z.~Tu, ``Convolutions and self-attention:
  {R}e-interpreting relative positions in pre-trained language models,''
  \emph{Annual Meeting of the Association for Computational Linguistics (ACL)},
  2021.

\bibitem{chen_simple_2021}
P.-C. Chen, H.~Tsai, S.~Bhojanapalli, H.~W. Chung, Y.-W. Chang, and C.-S.
  Ferng, ``A simple and effective positional encoding for transformers,''
  \emph{Empirical Methods in Natural Language Processing (EMNLP)}, 2021.

\bibitem{vaswani_attention_2017}
A.~Vaswani, N.~Shazeer, N.~Parmar, J.~Uszkoreit, L.~Jones, A.~N. Gomez,
  L.~Kaiser, and I.~Polosukhin, ``{Attention} {Is} {All} {You} {Need},''
  \emph{Annual Conference on Neural Information Processing Systems (NeurIPS)},
  2017.

\bibitem{choromanski_rethinking_2021}
K.~M. Choromanski, V.~Likhosherstov, D.~Dohan, X.~Song, A.~Gane, T.~Sarlos,
  P.~Hawkins, J.~Q. Davis, A.~Mohiuddin, L.~Kaiser, D.~B. Belanger, L.~J.
  Colwell, and A.~Weller, ``Rethinking {A}ttention with {P}erformers,''
  \emph{International Conference on Machine Learning (ICML)}, 2021.

\bibitem{katharopoulos_transformers_2020}
A.~Katharopoulos, A.~Vyas, N.~Pappas, and F.~Fleuret, ``Transformers are
  {RNN}s: {F}ast {A}utoregressive {T}ransformers with {L}inear {A}ttention,''
  \emph{International Conference on Machine Learning (ICML)}, 2020.

\bibitem{zhuoran_efficient_2021}
S.~Zhuoran, Z.~Mingyuan, Z.~Haiyu, Y.~Shuai, and L.~Hongsheng, ``Efficient
  {A}ttention: {A}ttention with {L}inear {C}omplexities,'' \emph{IEEE Winter
  Conference on Applications of Computer Vision (WACV)}, 2021.

\bibitem{shaw_rpe_2018}
P.~Shaw, J.~Uszkoreit, and A.~Vaswani, ``Self-attention with relative position
  representations,'' \emph{Annual Conference of the North American Chapter of
  the Association for Computational Linguistics (NAACL)}, 2018.

\bibitem{black_gptneox_2022}
S.~Black, S.~Biderman, E.~Hallahan, Q.~Anthony, L.~Gao, L.~Golding, H.~He,
  C.~Leahy, K.~McDonell, J.~Phang, M.~Pieler, U.~S. Prashanth, S.~Purohit,
  L.~Reynolds, J.~Tow, B.~Wang, and S.~Weinbach, ``{GPT}-{N}eo{X}-20{B}: An
  open-source autoregressive language model,'' \emph{Annual Meeting of the
  Association for Computational Linguistics (ACL)}, 2022.

\bibitem{gemma_gemma_2024}
\BIBentryALTinterwordspacing
G.~Team, T.~Mesnard, C.~Hardin, R.~Dadashi, S.~Bhupatiraju, S.~Pathak,
  L.~Sifre, M.~Rivière, M.~S. Kale, J.~Love, P.~Tafti, L.~Hussenot, P.~G.
  Sessa, A.~Chowdhery, A.~Roberts, A.~Barua, A.~Botev, A.~Castro-Ros, A.~Slone,
  A.~Héliou, A.~Tacchetti, A.~Bulanova, A.~Paterson, B.~Tsai, B.~Shahriari,
  C.~L. Lan, C.~A. Choquette-Choo, C.~Crepy, D.~Cer, D.~Ippolito, D.~Reid,
  E.~Buchatskaya, E.~Ni, E.~Noland, G.~Yan, G.~Tucker, G.-C. Muraru,
  G.~Rozhdestvenskiy, H.~Michalewski, I.~Tenney, I.~Grishchenko, J.~Austin,
  J.~Keeling, J.~Labanowski, J.-B. Lespiau, J.~Stanway, J.~Brennan, J.~Chen,
  J.~Ferret, J.~Chiu, J.~Mao-Jones, K.~Lee, K.~Yu, K.~Millican, L.~L. Sjoesund,
  L.~Lee, L.~Dixon, M.~Reid, M.~Mikuła, M.~Wirth, M.~Sharman, N.~Chinaev,
  N.~Thain, O.~Bachem, O.~Chang, O.~Wahltinez, P.~Bailey, P.~Michel, P.~Yotov,
  R.~Chaabouni, R.~Comanescu, R.~Jana, R.~Anil, R.~McIlroy, R.~Liu, R.~Mullins,
  S.~L. Smith, S.~Borgeaud, S.~Girgin, S.~Douglas, S.~Pandya, S.~Shakeri,
  S.~De, T.~Klimenko, T.~Hennigan, V.~Feinberg, W.~Stokowiec, Y.~hui Chen,
  Z.~Ahmed, Z.~Gong, T.~Warkentin, L.~Peran, M.~Giang, C.~Farabet, O.~Vinyals,
  J.~Dean, K.~Kavukcuoglu, D.~Hassabis, Z.~Ghahramani, D.~Eck, J.~Barral,
  F.~Pereira, E.~Collins, A.~Joulin, N.~Fiedel, E.~Senter, A.~Andreev, and
  K.~Kenealy, ``Gemma: Open models based on gemini research and technology,''
  2024. [Online]. Available: \url{https://arxiv.org/abs/2403.08295}
\BIBentrySTDinterwordspacing

\bibitem{touvron_llama_2023}
\BIBentryALTinterwordspacing
H.~Touvron, T.~Lavril, G.~Izacard, X.~Martinet, M.-A. Lachaux, T.~Lacroix,
  B.~Rozière, N.~Goyal, E.~Hambro, F.~Azhar, A.~Rodriguez, A.~Joulin,
  E.~Grave, and G.~Lample, ``Llama: Open and efficient foundation language
  models,'' 2023. [Online]. Available: \url{https://arxiv.org/abs/2302.13971}
\BIBentrySTDinterwordspacing

\bibitem{chowdhery_palm_2023}
A.~Chowdhery, S.~Narang, J.~Devlin, M.~Bosma, G.~Mishra, A.~Roberts, P.~Barham,
  H.~W. Chung, C.~Sutton, S.~Gehrmann, P.~Schuh, K.~Shi, S.~Tsvyashchenko,
  J.~Maynez, A.~Rao, P.~Barnes, Y.~Tay, N.~Shazeer, V.~Prabhakaran, E.~Reif,
  N.~Du, B.~Hutchinson, R.~Pope, J.~Bradbury, J.~Austin, M.~Isard, G.~Gur-Ari,
  P.~Yin, T.~Duke, A.~Levskaya, S.~Ghemawat, S.~Dev, H.~Michalewski, X.~Garcia,
  V.~Misra, K.~Robinson, L.~Fedus, D.~Zhou, D.~Ippolito, D.~Luan, H.~Lim,
  B.~Zoph, A.~Spiridonov, R.~Sepassi, D.~Dohan, S.~Agrawal, M.~Omernick, A.~M.
  Dai, T.~S. Pillai, M.~Pellat, A.~Lewkowycz, E.~Moreira, R.~Child, O.~Polozov,
  K.~Lee, Z.~Zhou, X.~Wang, B.~Saeta, M.~Diaz, O.~Firat, M.~Catasta, J.~Wei,
  K.~Meier-Hellstern, D.~Eck, J.~Dean, S.~Petrov, and N.~Fiedel, ``Palm:
  scaling language modeling with pathways,'' \emph{J. Mach. Learn. Res.},
  vol.~24, no.~1, Jan. 2023.

\bibitem{kaikondev_things_2023}
\BIBentryALTinterwordspacing
kaikondev, ``Things i'm learning while training superhot,'' 2023. [Online].
  Available: \url{https://kaiokendev.github.io/til}
\BIBentrySTDinterwordspacing

\bibitem{chen_extending_2023}
\BIBentryALTinterwordspacing
S.~Chen, S.~Wong, L.~Chen, and Y.~Tian, ``Extending context window of large
  language models via positional interpolation,'' \emph{arXiv}, 2023. [Online].
  Available: \url{https://arxiv.org/abs/2306.15595}
\BIBentrySTDinterwordspacing

\bibitem{wang_resonance_2024}
S.~Wang, I.~Kobyzev, P.~Lu, M.~Rezagholizadeh, and B.~Liu, ``Resonance rope:
  Improving context length generalization of large language models,''
  \emph{Annual Meeting of the Association for Computational Linguistics (ACL)},
  2024.

\bibitem{pawar_what_2024}
\BIBentryALTinterwordspacing
S.~Pawar, S.~M. T.~I. Tonmoy, S.~M.~M. Zaman, V.~Jain, A.~Chadha, and A.~Das,
  ``The what, why, and how of context length extension techniques in large
  language models -- a detailed survey,'' \emph{arXiv}, 2024. [Online].
  Available: \url{https://arxiv.org/abs/2401.07872}
\BIBentrySTDinterwordspacing

\bibitem{barbero_round_2025}
F.~Barbero, A.~Vitvitskyi, C.~Perivolaropoulos, R.~Pascanu, and
  P.~Veličković, ``Round and round we go! what makes rotary positional
  encodings useful?'' \emph{International Conference on Learning
  Representations (ICLR)}, 2025.

\bibitem{scholkopf_learning_2001}
B.~Scholkopf and A.~J. Smola, \emph{Learning with Kernels: Support Vector
  Machines, Regularization, Optimization, and Beyond}.\hskip 1em plus 0.5em
  minus 0.4em\relax MIT Press, 2001.

\bibitem{rahimi_weighted_2008}
A.~Rahimi and B.~Recht, ``Weighted sums of random kitchen sinks: replacing
  minimization with randomization in learning,'' \emph{Annual Conference on
  Neural Information Processing Systems (NeurIPS)}, 2008.

\bibitem{qin_linearized_2023}
Z.~Qin, W.~Sun, K.~Lu, H.~Deng, D.~Li, X.~Han, Y.~Dai, L.~Kong, and Y.~Zhong,
  ``Linearized relative positional encoding,'' \emph{Transactions on Machine
  Learning Research}, 2023.

\bibitem{wang_pop909_2020}
Z.~Wang, K.~Chen, J.~Jiang, Y.~Zhang, M.~Xu, S.~Dai, X.~Gu, and G.~Xia,
  ``{POP909}: {A} {Pop}-song {Dataset} for {Music} {Arrangement}
  {Generation},'' \emph{Conference of the International Society for Music
  Information Retrieval (ISMIR)}, 2020.

\bibitem{dai_automatic_2020}
S.~Dai, H.~Zhang, and R.~B. Dannenberg, ``Automatic {Analysis} and {Influence}
  of {Hierarchical} {Structure} on {Melody}, {Rhythm} and {Harmony} in
  {Popular} {Music},'' \emph{Joint Conference on AI Music Creativity (AIMC)},
  2020.

\bibitem{bengio_curriculum_2009}
Y.~Bengio, J.~Louradour, R.~Collobert, and J.~Weston, ``{Curriculum
  learning},'' \emph{International Conference on Machine Learning (ICML)},
  2009.

\bibitem{wang_supervised_2022}
J.-C. Wang, J.~B.~L. Smith, W.-T. Lu, and X.~Song, ``Supervised metric learning
  for music structure features,'' \emph{Conference of the International Society
  for Music Information Retrieval (ISMIR)}, 2021.

\bibitem{wu_musemorphose_2021}
S.-L. Wu and Y.-H. Yang, ``Musemorphose: Full-song and fine-grained piano music
  style transfer with one transformer vae,'' \emph{IEEE/ACM Transactions on
  Audio, Speech, \& Language Processing (TASLP)}, vol.~31, pp. 1953--1967,
  2023.

\bibitem{haki_real_2022}
B.~Haki, M.~Nieto, T.~Pelinski, and S.~Jord{\`a}~Puig, ``{R}eal-{t}ime {d}rum
  {a}ccompaniment {u}sing {t}ransformer {a}rchitecture,'' \emph{Joint
  Conference on AI Music Creativity (AIMC)}, 2022.

\bibitem{moore_stats_1999}
D.~S. Moore and S.~Kirkland, \emph{The Basic Practice of Statistics},
  2nd~ed.\hskip 1em plus 0.5em minus 0.4em\relax W. H. Freeman \& Co., 1999.

\bibitem{levene_robust_1961}
H.~Levene, ``Robust tests for equality of variances,'' \emph{Contributions to
  probability and statistics. Essays in honor of Harold Hotelling}, pp.
  279--292, 1961.

\bibitem{welch_generalization_1947}
B.~L. Welch, ``The generalization of `student's' problem when several different
  population variances are involved,'' \emph{Biometrika}, vol.~34, no. 1/2, pp.
  28--35, 1947.

\bibitem{favero_compositional_2025}
\BIBentryALTinterwordspacing
A.~Favero, A.~Sclocchi, F.~Cagnetta, P.~Frossard, and M.~Wyart, ``How
  compositional generalization and creativity improve as diffusion models are
  trained,'' \emph{arXiv}, 2025. [Online]. Available:
  \url{https://arxiv.org/abs/2502.12089}
\BIBentrySTDinterwordspacing

\bibitem{mei_power_2025}
\BIBentryALTinterwordspacing
K.~Mei, H.~Talebi, M.~Ardakani, V.~M. Patel, P.~Milanfar, and M.~Delbracio,
  ``The power of context: How multimodality improves image super-resolution,''
  \emph{arXiv}, 2025. [Online]. Available:
  \url{https://arxiv.org/abs/2503.14503}
\BIBentrySTDinterwordspacing

\end{thebibliography}


\vfill

\end{document}